\documentclass[aps, prd, superscriptaddress, preprintnumbers, showpacs]{revtex4}
\usepackage{graphicx,amsfonts,amsmath,amssymb,amstext}
\usepackage{float,wrapfig}
\usepackage{subfigure, psfrag}
\usepackage{dsfont}
\usepackage{color}
\usepackage{bm}
\usepackage{multirow,textcomp}
\usepackage{enumerate}

\newcommand{\be}{\begin{equation}}
\newcommand{\ee}{\end{equation}}
\newcommand{\ba}{\begin{eqnarray}}
\newcommand{\ea}{\end{eqnarray}}

\newcommand{\sign}{\,\mbox{sign}}

\definecolor{purple}{rgb}{0.8,0,0.6}

\makeatletter
\newcommand{\vast}{\bBigg@{2}}
\newcommand{\Vast}{\bBigg@{3}}
\makeatother

\begin{document}

\title{Chiral separation and chiral magnetic effects in a slab: The role of boundaries}
\date{\today}

\author{E. V. Gorbar}
\affiliation{Department of Physics, Taras Shevchenko National Kiev University, Kiev, 03680, Ukraine}
\affiliation{Bogolyubov Institute for Theoretical Physics, Kiev, 03680, Ukraine}

\author{V. A. Miransky}
\affiliation{Department of Applied Mathematics, Western University, London, Ontario N6A 5B7, Canada}
\affiliation{Department of Physics and Astronomy, Western University, London, Ontario N6A 3K7, Canada}

\author{I. A. Shovkovy}
\affiliation{College of Letters and Sciences, Arizona State University, Mesa, Arizona 85212, USA}

\author{P. O. Sukhachov}
\affiliation{Department of Physics, Taras Shevchenko National Kiev University, Kiev, 03680, Ukraine}

\begin{abstract}
We study the chiral separation and chiral magnetic effects in a slab of Dirac semimetal of finite thickness,
placed in a constant magnetic field perpendicular to its surfaces. We utilize the Bogolyubov boundary
conditions with a large Dirac mass (band gap) outside the slab. We find that, in a finite thickness slab,
the axial current density is induced by helicity-correlated standing waves and, as a consequence, is
quantized. The quantization is seen in its stepped-shape dependence on the fermion chemical potential
and a sawtooth-shape dependence on the thickness of the slab. In contrast to a naive expectation,
there is no chiral charge accumulation anywhere in the bulk or at the boundaries of the semimetal.
In the same slab geometry, we also find that a nonzero chiral chemical potential induces no electric current,
as might have been expected from the chiral magnetic effect. We argue that this outcome is natural and
points to the truly non-static nature of the latter. By taking into account a nonzero electric field of
a double layer near the boundaries of the slab, we find that the low-energy modes under consideration
satisfy the continuity equation for axial current density without the anomalous term.
\end{abstract}


\maketitle

\section{Introduction}
\label{sec:Introduction}

Nowadays there is significant interest in relativistic matter in a strong magnetic field. Such matter is
intensively studied both experimentally and theoretically. It has a number of applications in high-energy
physics and astrophysics (e.g., in the context of compact stars, heavy-ion collisions, and the early universe),
as well as in condensed matter physics (e.g., in the context of novel Dirac/Weyl materials).

In condensed matter physics, after the first reports of three-dimensional Dirac semimetals appeared
two years ago \cite{Borisenko,Neupane,Liu}, the field exploded with numerous investigations
of their exotic properties (for reviews, see Refs.~\cite{Turner,Vafek,Burkov}). One of the key aspects of
such materials is a well-defined chirality of the low-energy quasiparticles, described by Weyl
fermions. Because of the chiral anomaly \cite{ABJ}, the chirality is not a conserved charge and, thus,
the corresponding quasiparticles must come in both chiralities \cite{Nielsen:1983rb}. In Dirac semimetals, the
low-energy spectra of the quasiparticles of opposite chiralities are degenerate. Such 
degeneracy is often protected by symmetries (e.g., the symmetry under time-reversal or parity).
If the corresponding symmetry is broken, however, a Dirac semimetal may turn into a Weyl
semimetal, in which the degeneracy of the quasiparticles with opposite chiralities is lifted.
A number of materials of this latter type has been recently reported as well
\cite{1501.00060,1501.00755,1502.00251,1502.03807,1502.04684,1503.01304,1504.01350}.

Phenomenologically, a chiral asymmetry in relativistic matter may be introduced via a nonzero
chiral chemical potential $\mu_5$ \cite{Fukushima:2008xe}. Such a chemical potential couples to a
difference between the number densities of the left- and right-handed fermions and enters the Lagrangian
density through the term $\mu_5\bar{\psi}\gamma^0\gamma^5\psi$, where $\bar{\psi}\equiv \psi^\dagger\gamma^0$.
The latter produces a chiral asymmetry
in magnetized relativistic matter and leads to a nondissipative electric current $\mathbf{j}=e^2\mathbf{B}
\mu_5/(2\pi^2)$ in the presence of an external magnetic field $\mathbf{B}$ \cite{Kharzeev:2007tn,
Kharzeev:2007jp,Fukushima:2008xe} (see also Refs.~\cite{BurkovBalents,BurkovHookBalents,
ZyuzinBurkov,Franz}). This phenomenon is known in the literature as the chiral
magnetic effect (CME) and its origin is related to the famous chiral anomaly \cite{ABJ}. Moreover,
the charge-dependent correlations and flow, observed in heavy-ion collisions at RHIC
\cite{collisions,Wang:2012qs,Ke:2012qb,Adamczyk:2013kcb,Adamczyk:2015eqo} and LHC \cite{Selyuzhenkov:2011xq},
appear to be in a qualitative agreement with the predictions of the CME \cite{Voloshin:2004vk,
Kharzeev:2009fn} (for recent reviews, see also Refs.~\cite{Liao:2014ava,Kharzeev:2015kna,Huang:2015oca}).
In the context of condensed matter physics, it was also suggested that the measured quadratic field
dependence of the magnetoconductance in $\mathrm{ZrTe_5}$ is an indication of the chiral magnetic
effect \cite{Li-Kharzeev:2014bha}.

Unlike the chiral chemical potential, which is a rather exotic quantity and is not very well defined
theoretically, the chemical potential $\mu$ (associated, for example, with a conserved electric or
baryon charge) is common in many physical systems. It was shown in Refs.~\cite{Vilenkin:1980ft,
Zhitnitsky,Newman} that a nondissipative axial current density $\mathbf{j}_5=-e\mathbf{B}\mu/(2\pi^2)$
exists in the equilibrium state of noninteracting massless fermion matter in a magnetic field. This
effect is known as the chiral separation effect (CSE). In fact, as suggested in Refs.~\cite{chiral-shift-2,
Burnier:2011bf}, the CSE may lead to a chiral charge separation (i.e., effectively inducing a
nonzero chiral chemical potential $\mu_5$) and, thus, trigger the CME even in the absence of
topological fluctuations in the initial state.

The physical and mathematical reasons for the chiral asymmetry in relativistic matter in a magnetic
field are quite transparent (for an elegant exposition, see also Ref.~\cite{Basar:2012gm}). In essence,
its origin is connected with the spin-polarized nature of the lowest Landau level (LLL). The corresponding
fermionic states are also characterized by a well-defined longitudinal momentum and, thus, chirality.
Moreover, the states with opposite signs of the longitudinal momenta carry opposite chiralities and,
thus, lead to a nondissipative axial current density $\mathbf{j}_5=-e\mathbf{B} \mu/(2\pi^2)$
\cite{Vilenkin:1980ft,Zhitnitsky}.

It was argued in Refs.~\cite{Zhitnitsky,Newman} that nondissipative currents in magnetized relativistic
matter are determined by the topological currents induced exclusively in the LLL and are intimately
connected with the chiral anomaly. This fact is directly connected with the well-known result that the
chiral anomaly in a magnetic field is also generated exclusively by the LLL \cite{Ambjorn}. The first
studies of interaction effects on the chiral asymmetry of relativistic matter in a magnetic field were
performed in Refs.~\cite{chiral-shift-1,Fukushima,chiral-shift-2} by using Nambu--Jona-Lasinio
models with local interaction. In particular, it was found that the interaction unavoidably generates
a chiral shift $\Delta$ \cite{chiral-shift-1,chiral-shift-2} when the fermion density is nonzero. It enters
the effective Lagrangian density through the following quadratic term: $\Delta \bar\psi \gamma^3
\gamma^5 \psi $, when the magnetic field is directed along the $z$ direction.
The meaning of the chiral shift parameter is most transparent in the chiral limit:
it determines a relative shift of the longitudinal momenta in the dispersion relations of opposite
chirality fermions, $k^{3}\to k^{3}\pm\Delta$, where the momentum $k^{3}$ is directed along the
magnetic field. Furthermore, as shown in Refs.~\cite{chiral-shift-1,chiral-shift-2,chiral-shift-3}, the
chiral shift $\Delta$ is responsible for an additional contribution to the axial current density.
Also, such a dynamically generated chiral shift splits each Dirac node
into a pair of Weyl nodes of opposite chirality, thus producing a Weyl semimetal
from a Dirac one \cite{chiral-shift-4}. Recently, another interesting mechanism
for inducing a nonzero chiral shift was proposed \cite{Ebihara:2015aca,Chan:2015dwa}.
It uses a circularly polarized light and works even at zero chemical potential.

Usually the chiral magnetic and chiral separation effects are considered in the literature in unbounded
material media. In practice, however, all physical systems (except for the CME and CSE in the early universe,
perhaps) are finite. It is natural then to ask about the role of boundaries and finite-size effects
in the chiral magnetic and separation effects. Indeed, even if one assumes that, in the bulk of a bounded
medium, the electric and axial currents are the same as those in infinite systems, they should get modified
near the boundaries. This should be an immediate consequence of the continuity equations if the currents
are required to vanish outside the material. This simple observation was the main motivation for the present
work. Here we will study the chiral separation and chiral magnetic effects in a slab with an external magnetic
field perpendicular to the boundary planes of the slab, which yields the simplest realization of a finite-size
system.

The paper is organized as follows. In Sec.~\ref{sec:Model} we introduce the model of a slab with the
Bogolyubov boundary conditions \cite{Bogoliubov}, when the mass (band gap) parameter in the
vacuum outside the slab is taken to be the largest mass (energy) parameter in the model. Note
that we will use both the terms ``mass" and ``band gap" interchangeably throughout the paper.
Section~\ref{sec:AC-slab} is devoted to the analysis of the axial current density (i.e., the CSE) in the slab with such boundary
conditions. The CME is considered in Sec.~\ref{sec:chiral-current-mu5}. In Sec.~\ref{sec:electric-field}
we discuss the chiral anomaly in vacuum regions near the surface of the slab. The discussion of the main
results is given in Sec.~\ref{sec:Conclusion}. Some technical details, including the derivation of the
Landau-level wave functions and the implementation of the boundary conditions, are presented in
Appendices~\ref{App:WF} and \ref{App:BC}, respectively.

Throughout this paper, we set $\hbar=1$ and $c=1$.

\section{Model}
\label{sec:Model}

The Hamiltonian of the low-energy model of a Dirac semimetal slab situated between the planes
$z=-a$ and $z=a$ reads
\begin{equation}
H= \int d^3 \mathbf{r} \, \Psi^\dagger (\mathbf{r})
\left[v_F \bm{\alpha} \cdot(-i\bm{\nabla}+e\mathbf{A}) + \gamma^0 m(z) \right]
\Psi (\mathbf{r}),
\label{hamiltonian-0}
\end{equation}
where $\bm{\alpha} = \gamma^0\bm{\gamma}$, and $\bm{\gamma}$ are the Dirac matrices in the
chiral representation, i.e.,
\begin{equation}
\gamma^0 = \left( \begin{array}{cc} 0 & -I_{2}\\ -I_{2} & 0 \end{array} \right),\qquad
\bm{\gamma} = \left( \begin{array}{cc} 0& \bm{\sigma} \\  - \bm{\sigma} & 0 \end{array} \right), \qquad \gamma^5 \equiv
i\gamma^0\gamma^1\gamma^2\gamma^3 = \left( \begin{array}{cc} I_{2} & 0\\ 0 & -I_{2} \end{array} \right).
\label{Dirac-matrices}
\end{equation}
Here $I_{2}$ is the two-dimensional unit matrix and $\bm{\sigma}=(\sigma_1,\sigma_2,\sigma_3)$
are the Pauli matrices. By assuming that the external magnetic field $\mathbf{B}$ is directed along the $z$
axis, we will find it convenient to use the vector potential in the Landau gauge, i.e., $\mathbf{A}=(0, Bx, 0)$.
The other notations are as follows: $e$ is the electron charge, $v_F$ is  the Fermi velocity, and
$m(z)=M\, \theta(z^2-a^2)+m\, \theta(a^2-z^2)$ is the Dirac mass (band gap) function [here $\theta(x)$
is the unit step function]. The case of a Dirac semimetal with a zero band gap in the bulk is
obtained by taking the limit $m\to 0$. In the model at hand, we assume that the ``vacuum" gap parameter
$M$ is much larger than all characteristic energy scales in the slab (e.g., the work function and/or
relevant quasiparticle energies). Interestingly, such a model of the slab is nothing else but a generalized
Bogolyubov bag model \cite{Bogoliubov}. In the studies of graphene, similar boundary conditions with
an infinite gap outside the material are known as the infinite mass boundary conditions \cite{Beenakker,Recher}.
The same idea, albeit with a finite-size band gap, is also utilized in modeling a potential barrier
in the context of the Klein paradox in graphene; see Ref.~\cite{Katsnelson}. While $m=0$ was 
used in the original Bogolyubov model, we will treat $m$ as a free parameter in the analysis below. 
(Note that the $m\neq 0$ case is of interest not only from a theoretical viewpoint, but could be also 
investigated experimentally in a more general class of Dirac semimetals/metals, e.g., such as a 
bismuth alloy Bi$_{1-x}$Sb$_x$ at small concentrations of antimony \cite{Lenoir,Teo}, where the 
Dirac gap is nonzero.) For a good review of bag models in hadron physics, see Ref.~\cite{Thomas}.
Generically, in all such models, the hadrons are described as bags with massless fermions (quarks)
confined inside. In order to prevent massless fermions from leaving the bag, one requires that
the normal components of the momenta and hence the currents across the surface vanish. Such
boundary conditions necessarily break the chiral symmetry for fermions \cite{Thomas}. From the physics viewpoint, this
is unavoidable because massless quasiparticles experience a helicity flip (and, thus, a chirality change)
whenever the directions of their momenta change due to scattering from the boundary. This fact will
be crucial for our analysis below. In particular, in the Bogolyubov model with the vanishing gap
in the bulk, $m\to 0$, the chiral symmetry is explicitly broken by the inclusion of the (infinitely) large
vacuum gap parameter $M$.

\section{Chiral separation effect}
\label{sec:AC-slab}

In this section, we calculate the axial current density and study the CSE in a Dirac semimetal with a
slab geometry. By making use of Hamiltonian (\ref{hamiltonian-0}), the ground state of the
system will be obtained by filling all quasiparticle states with the energies less than the Fermi
energy $E_F = \mu$, where $\mu$ is the chemical potential.

Let us start by determining the energy spectrum and the electron wave functions for a Dirac
semimetal slab in a constant magnetic field. Before proceeding to the slab case, however, it
is instructive to start from presenting the Landau-level wave functions in an infinite space.
By making use of the chiral representation (\ref{Dirac-matrices}), we derive the following
results for the wave functions (see Appendix~\ref{App:WF} for details):
\begin{eqnarray}
\label{LLL}
\psi(\mathbf{r})_{n=0} &=&C_0\,e^{ip_zz+ip_yy}\left(
 \begin{array}{c}
   0 \\
    Y_{0}(\xi) \\
    0  \\
    -\frac{m}{E_0-v_Fp_z}Y_{0}(\xi) \\
 \end{array}
   \right),\\
\label{higher}
\psi(\mathbf{r})_{n > 0}&=&e^{ip_zz+ip_yy} \left[C_{+}\left(
 \begin{array}{c}
   -i\frac{m^2+2n\epsilon_{L}^2}{(E_n-v_F p_z)\sqrt{2n\epsilon_{L}^2}} Y_{n-1}(\xi) \\
    Y_{n}(\xi) \\
    i\frac{m}{\sqrt{2n\epsilon_{L}^2}} Y_{n-1}(\xi)  \\
    0 \\
 \end{array}
   \right) +C_{-}\left(
 \begin{array}{c}
   -i\frac{m\sqrt{m^2+2n\epsilon_{L}^2}}{(E_n-v_F p_z)\sqrt{2n\epsilon_{L}^2}} Y_{n-1}(\xi) \\
    0 \\
    i\frac{\sqrt{m^2+2n\epsilon_{L}^2}}{\sqrt{2n\epsilon_{L}^2}} Y_{n-1}(\xi)  \\
    \frac{\sqrt{m^2+2n\epsilon_{L}^2}}{E_n-v_F p_z} Y_{n}(\xi) \\
 \end{array}
   \right)\right],
\end{eqnarray}
where $l=1/\sqrt{|eB|}$ is the magnetic length, $\epsilon_{L} = v_F\sqrt{eB}$ is the Landau energy scale, and
$\xi=x/l+p_y l$. For convenience, here, we fixed the sign of electric charge so that $\sign{(eB)}=+1$. Additionally,
we introduced the following harmonic oscillator wave functions:
$Y_n(\xi)=\frac{e^{-\xi^2/2}}{\sqrt{2^n n!\sqrt{\pi}} }H_n(\xi)$,
where $H_n(\xi)$ are the Hermite polynomials. (For the Landau-level wave functions in the standard
representation of the Dirac matrices, see Ref.~\cite{wf}.) The corresponding Landau-level energies are
given by $E_n=\pm\sqrt{v_F^2p_z^2+m^2+2n\epsilon_{L}^2}$.

Now, in the case of a slab with a finite extent in the $z$ direction, for every plane wave with a wave vectoror
$p_z$, propagating in the positive $z$ direction, there should be also a plane wave with a wave vector
$-p_z$, propagating in the opposite direction. Therefore, the general solution for the $n$th-Landau-level
wave function in the slab is given by a superposition of two counterpropagating plane waves,
or equivalently, a standing wave:
\begin{eqnarray}
\Psi_{\rm slab}(\mathbf{r})_{n=0} &=&C_0\,e^{ip_zz+ip_yy}\left(
 \begin{array}{c}
   0 \\
    Y_{0}(\xi) \\
    0  \\
    -\frac{m}{E_0-v_Fp_z}Y_{0}(\xi) \\
 \end{array}
   \right) + \tilde{C}_0\,e^{-ip_zz+ip_yy}\left(
 \begin{array}{c}
   0 \\
    Y_{0}(\xi) \\
    0  \\
    -\frac{m}{E_0+v_Fp_z}Y_{0}(\xi) \\
 \end{array}
   \right),
   \label{LLL-slab}
\\
\Psi_{\rm slab}(\mathbf{r})_{n > 0}&=&e^{ip_zz+ip_yy} \left[C_{+}\left(
 \begin{array}{c}
   -i\frac{m^2+2n\epsilon_{L}^2}{(E_n-v_F p_z)\sqrt{2n\epsilon_{L}^2}} Y_{n-1}(\xi) \\
    Y_{n}(\xi) \\
    i\frac{m}{\sqrt{2n\epsilon_{L}^2}} Y_{n-1}(\xi)  \\
    0 \\
 \end{array}
   \right) +C_{-}\left(
 \begin{array}{c}
   -i\frac{m\sqrt{m^2+2n\epsilon_{L}^2}}{(E_n-v_F p_z)\sqrt{2n\epsilon_{L}^2}} Y_{n-1}(\xi) \\
    0 \\
    i\frac{\sqrt{m^2+2n\epsilon_{L}^2}}{\sqrt{2n\epsilon_{L}^2}} Y_{n-1}(\xi)  \\
    \frac{\sqrt{m^2+2n\epsilon_{L}^2}}{E_n-v_F p_z} Y_{n}(\xi) \\
 \end{array}
   \right)\right] \nonumber \\
                        &+& \left(p_z \to -p_z, C_{\pm} \to \tilde{C}_{\pm}\right).
                        \label{higher-slab}
\end{eqnarray}
These wave functions inside the slab should be matched to the corresponding solutions in the vacuum.
This is done in Appendix~\ref{App:BC} using the Bogolyubov bag model, in which wave functions
outside the semimetal satisfy the Dirac equation with an infinitely large vacuum gap parameter $M$.
By enforcing the boundary conditions, we also find that the wave vector $p_z$ should satisfy the following
spectral equation:
\begin{equation}
v_F p_z\cos{(2ap_z)}+m\sin{(2ap_z)}=0,
\label{spectrum-n0-matching-main}
\end{equation}
where $p_z\neq0$. The final expressions for the Landau-level wave functions in the slab
are given in Eqs.~(\ref{psi-matching-n0-slab-01}), (\ref{psi-n-all-InfMass1}), and (\ref{psi-n-all-InfMass2}).
In essence, they have the form of standing waves with discrete wave vectors $p_z$ that satisfy
Eq.~(\ref{spectrum-n0-matching-main}). It is important to note that, while there are two independent
solutions for each wave vector in the higher Landau levels, there is only one independent solution
for each wave vector in the LLL.

\subsection{Axial current density}

In this subsection, we calculate the axial current density inside the slab. In terms of the Landau-level wave
functions, the corresponding ground state expectation value is given by
\begin{eqnarray}
\langle j_{5}^{3}\rangle= \int \frac{dp_y}{2\pi} \sum_{p_z} \left( f(p_z)\, v_F\Psi_{\rm slab}^{\dag}(\mathbf{r})_{n=0}\gamma^0\gamma^3\gamma^5\Psi_{\rm slab}(\mathbf{r})_{n=0} +\sum^2_{i=1}\sum_{n=1}^{\infty} f(p_z)\,v_F\Psi_{\rm slab}^{(i)\,\dag}(\mathbf{r})_{n}\gamma^0\gamma^3\gamma^5\Psi^{(i)}_{\rm slab}(\mathbf{r})_{n} \right),
\label{axial-current-slab}
\end{eqnarray}
where the contributions of quasiparticles from both valence and conduction bands are taken into account
via the use of the following generalized distribution function:
\begin{equation}
f(p_z)=\frac{1}{e^{\left(\sqrt{v_F^2p_z^2+m^2+2n\epsilon_{L}^2}-\mu\right)/T}+1}-\frac{1}{e^{\left(\sqrt{v_F^2p_z^2+m^2+2n\epsilon_{L}^2}+\mu\right)/T}+1}.
\label{distribution}
\end{equation}
Here $\mu$ is the chemical potential (Fermi energy) measured from the Dirac point. This
distribution function accounts for the fact that the quasiparticles of the conduction and valence bands
carry opposite charges. In the zero-temperature limit, which we use in the following, this function
simplifies:
\begin{equation}
\lim_{T\to 0}f(p_z) =\sign{(\mu)}\theta\left(\mu^2-v_F^2p_z^2-m^2-2n\epsilon_{L}^2\right),
\label{distribution-T0}
\end{equation}
where $\theta(x)$ is the unit step function.

When calculating the axial current density, it is instructive to separate the contribution of the spin-polarized
lowest Landau level from the contributions of the higher Landau levels ($n>0$). The LLL contribution is obtained
by making use of the wave function in Eq.~(\ref{psi-matching-n0-slab-01}). The zero-temperature result reads
\begin{eqnarray}
\langle j_{5}^{3}\rangle_{n=0}&=&\int \frac{dp_y}{2\pi} \sum_{p_z} f(p_z)\,v_F \Psi_{\rm slab}^{\dag}
(\mathbf{r})_{n=0}\gamma^0\gamma^3\gamma^5\Psi_{\rm slab}(\mathbf{r})_{n=0} \nonumber \\
&=&- \frac{eBv_F\sign{(\mu)}}{2a\pi}\sum_{p_z} \theta\left(\mu^2-v_F^2p_z^2-m^2\right)
\frac{ \left(m^2+v_F^2p_z^2\right)\left[1-\cos{(2p_z z)}\cos{(2p_z a)}\right] }{2(m^2+v_F^2p_z^2)+ m v_F/a },
\label{axial-current-slab-n0}
\end{eqnarray}
where we also made use of the spectral equation in Eq.~(\ref{spectrum-n0-matching-main}). As we see,
in a general case when $m\neq 0$, the LLL contribution to the axial current density has a nontrivial
dependence on the $z$ coordinate. Here it may be appropriate to mention that the axial
current density is well defined even in the gapped case when the axial charge is not conserved.
In fact, even in the extreme non-relativistic limit, it has a clear physical meaning as a spin polarization.

As expected, in the chiral (gapless) limit, the axial current density is independent of the
$z$ coordinate and the explicit result reads
\begin{equation}
\langle j_{5}^{3}\rangle_{n=0,m\to 0} = - \frac{eBv_F\sign{(\mu)}}{4 a \pi}\sum_{p_z} \theta\left(\mu^2-v_F^2p_z^2\right)
= - \frac{eBv_F\sign{(\mu)}}{4 a \pi} k_{\rm max} \, ,
\label{axial-current-slab-n00}
\end{equation}
where we took into account the spectral equation (\ref{spectrum-n0-matching-main}), which reduces
down to $\cos{(2ap_z)}=0$ in the gapless case. The latter also implies that the allowed values of the wave vector
are $p^{(0)}_{z,k} = (2k-1) \pi /(4a)$, where $k$ is a positive integer. Because of the unit step function in
Eq.~(\ref{axial-current-slab-n00}), the result of the sum is given by $k_{\rm max}  = \left[2a|\mu|/(v_F\pi) +1/2\right]$
where $[\ldots]$ represents the integer part. As expected, in the limit of $a\to \infty$, the above result reduces to
the well-known relation for the chiral separation effect in an infinite system, i.e., $\langle j_{5}^{3}\rangle = - |eB| \mu/(2\pi^2)$.

The result in Eq.~(\ref{axial-current-slab-n00}) shows that, in a slab of finite thickness $a$, the axial
current density is quantized. This is a qualitatively new feature that did not exist in an infinite system. It is
a natural outcome of having the wave functions in the form of standing waves and the quantization
of the wave vector in a slab geometry. As is easy to see from Eq.~(\ref{axial-current-slab-n00}),
the height of the steps in the axial current is proportional to the magnetic field and inversely proportional
to the thickness of the slab, i.e., $\delta\langle j_{5}^{3}\rangle= eB v_F /(4 a \pi)$. When $\langle j_{5}^{3}\rangle$
is plotted as a function of $a\mu/v_F$, the widths of the steps are given by $\pi/2$.
It would be interesting to explore whether such a quantization can be observed in experiment.

In the case of higher Landau levels ($n>0$), there are separate contributions to the axial current density from
each of the two independent modes in the slab, i.e., $\Psi^{(1)}_{\rm slab}(\mathbf{r})_{n}$ and
$\Psi^{(2)}_{\rm slab}(\mathbf{r})_{n}$. By making use of the explicit expressions for the corresponding
wave functions in Eqs.~(\ref{psi-n-all-InfMass1}) and (\ref{psi-n-all-InfMass2}), we find that, for each $p_z$,
the two contributions have opposite signs, i.e.,
\begin{eqnarray}
\label{axial-current-slab-n-ij-InfMass-1}
\int \frac{dp_y}{2\pi}
v_F\Psi_{\rm slab}^{(1)\,\dag}(\mathbf{r})_{n}\gamma^0\gamma^3\gamma^5\Psi^{(1)}_{\rm slab}(\mathbf{r})_{n}
&=& -\frac{eB}{2\pi} \frac{v_Fp_z \left(m^2+v_F^2p_z^2\right)\left[1-\cos{(2p_z z)}\cos{(2p_z a)}\right] }
{2ap_zE_n^2+v_Fp_z m - n\epsilon_{L}^2 \sin{(4ap_z)}}, \\
\label{axial-current-slab-n-ij-InfMass-2}
\int \frac{dp_y}{2\pi}
v_F\Psi_{\rm slab}^{(2)\,\dag}(\mathbf{r})_{n}\gamma^0\gamma^3\gamma^5\Psi^{(2)}_{\rm slab}(\mathbf{r})_{n}
&=& \frac{eB}{2\pi} \frac{v_Fp_z\left(m^2+v_F^2p_z^2\right)\left[1-\cos{(2p_z z)}\cos{(2p_z a)}\right] }
{2ap_zE_n^2+v_Fp_zm - n\epsilon_{L}^2 \sin{(4ap_z)}}.
\end{eqnarray}
Therefore, there is no net contribution to the axial current density due to higher Landau levels.
In other words, just like in the case of an infinite space \cite{Zhitnitsky}, the axial current density in a
semimetal slab is determined exclusively by the LLL contribution (\ref{axial-current-slab-n0}).

\begin{figure*}[ht]
\begin{center}
\includegraphics[width=0.32\textwidth]{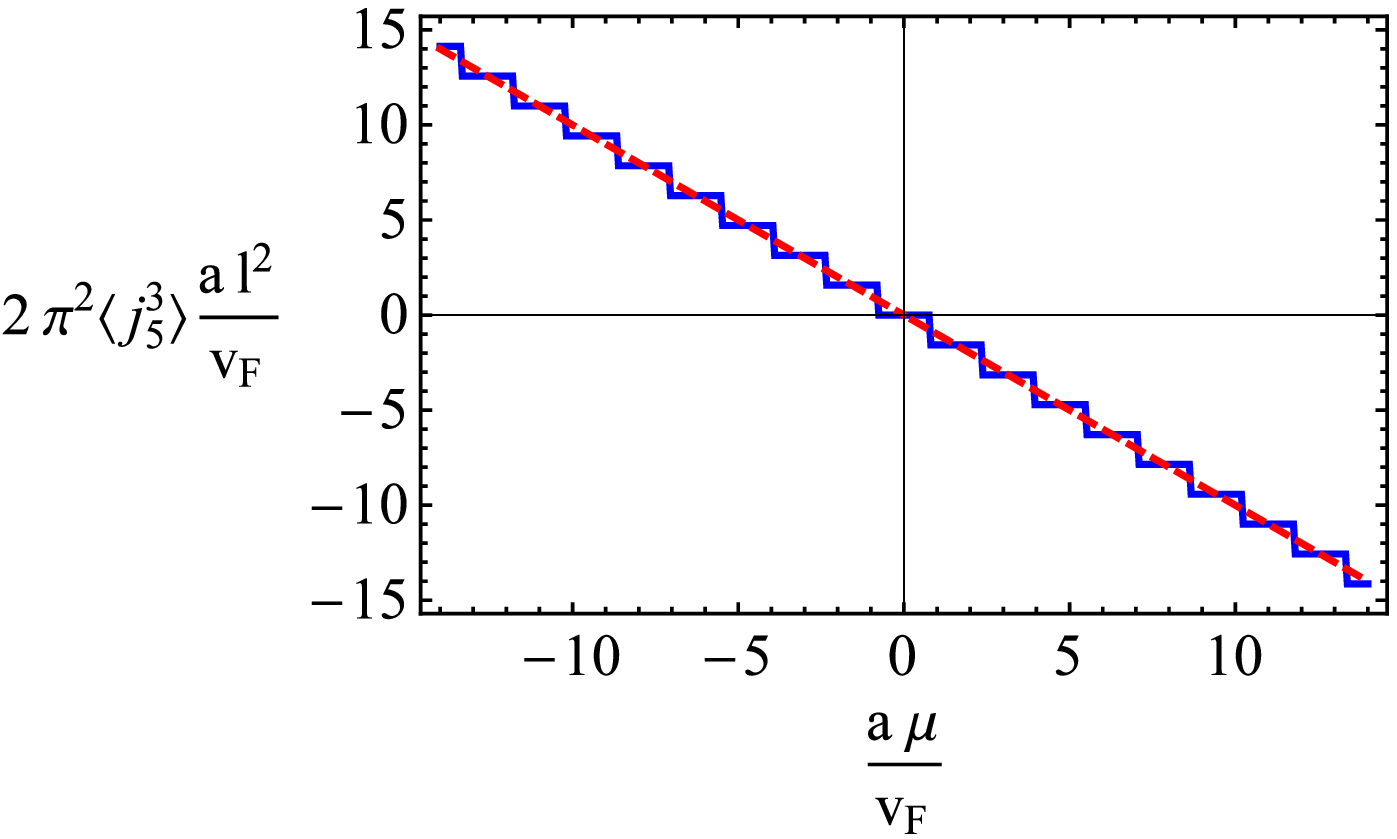}
\hfill
\includegraphics[width=0.32\textwidth]{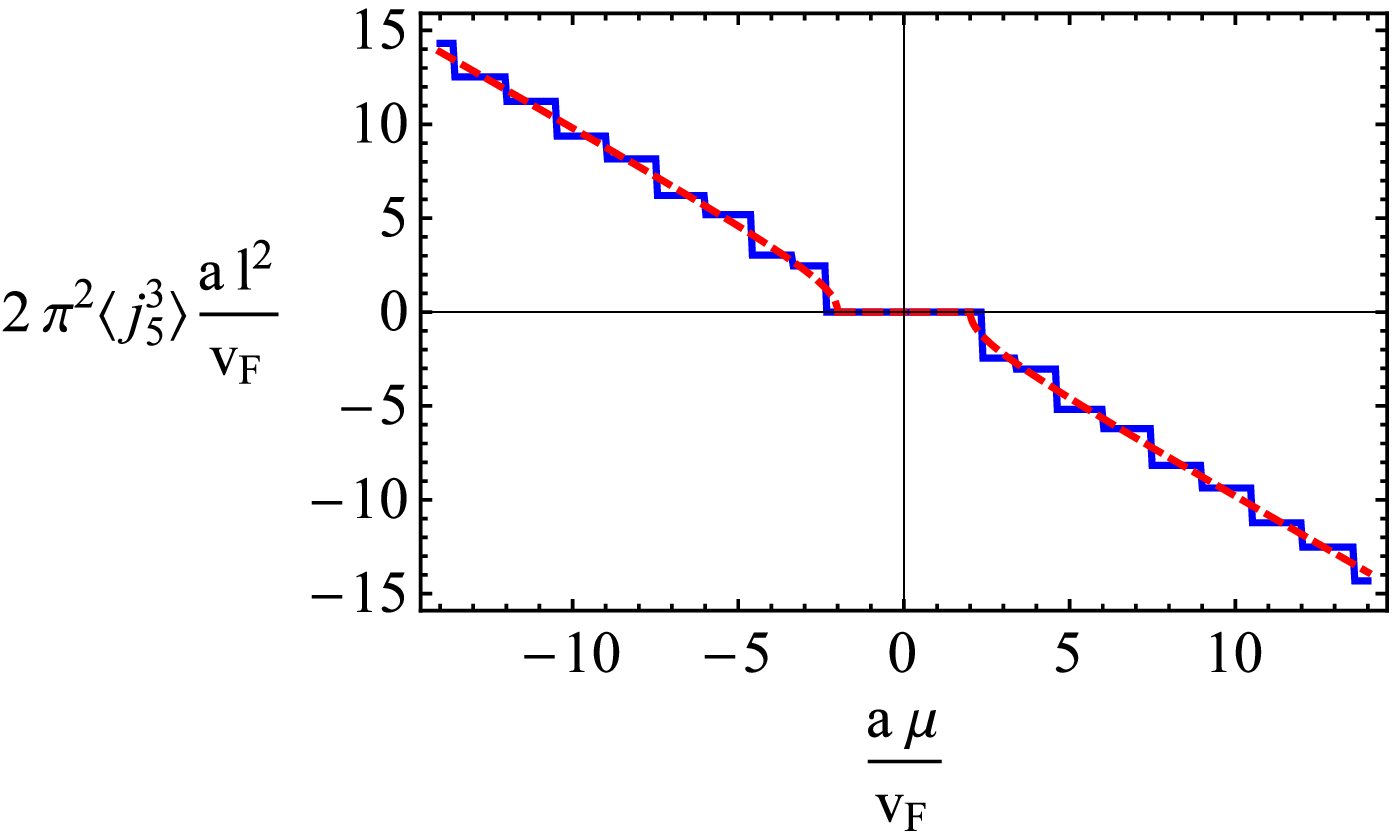}
\hfill
\includegraphics[width=0.32\textwidth]{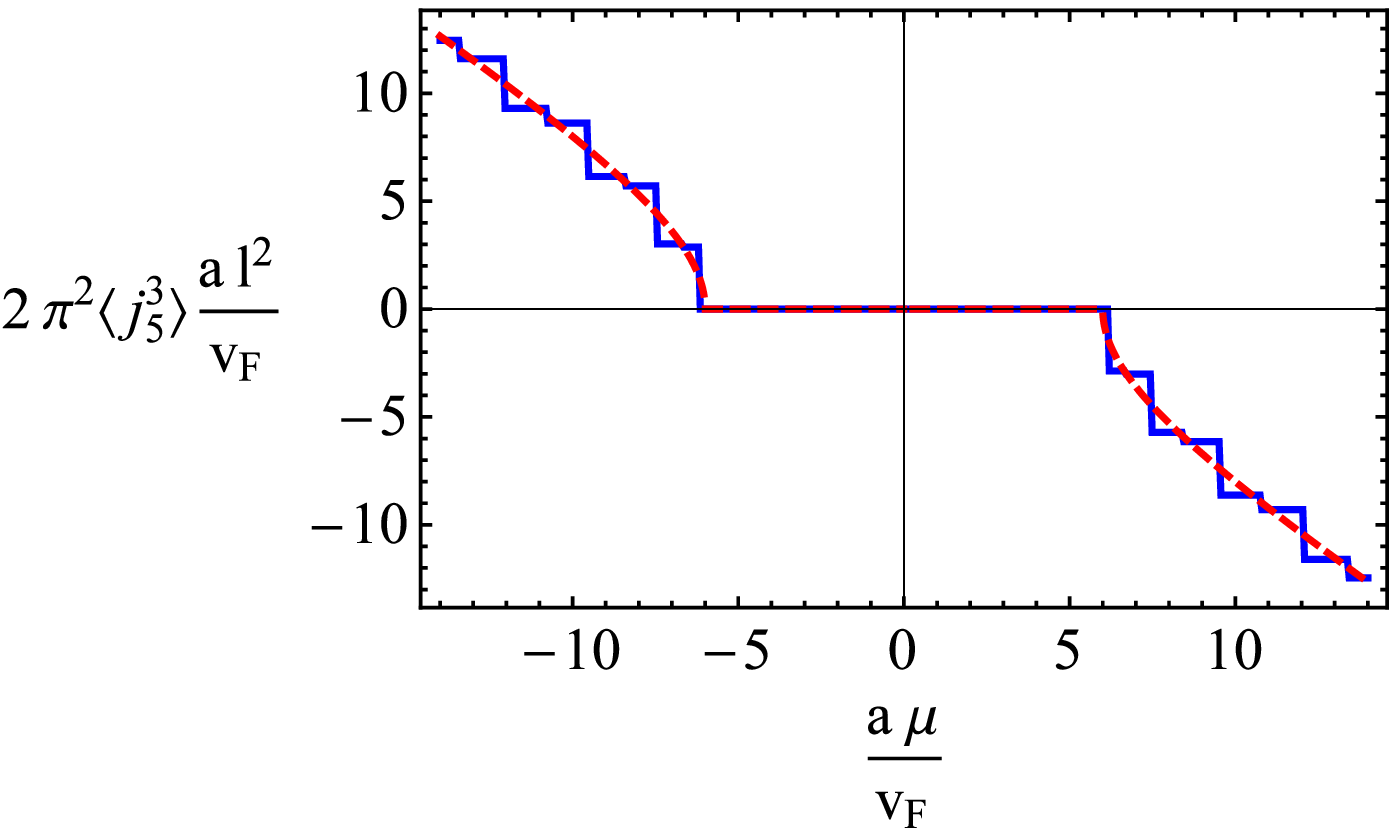}
\end{center}
\caption{(Color online) The dimensionless axial current density in an infinite space (red dashed line) and in the 
middle of the slab (blue solid line) plotted as a function of chemical potential for the three values of the band gap:
$am/v_F=0$ (left panel), $am/v_F=2$ (middle panel), and $am/v_F=6$ (right panel). To plot the figure, we 
fixed $v_F/a=25~\mbox{meV}$.}
\label{fig:axial_current_slab_n0_mu}
\end{figure*}

\begin{figure*}[ht]
\begin{center}
\includegraphics[width=0.32\textwidth]{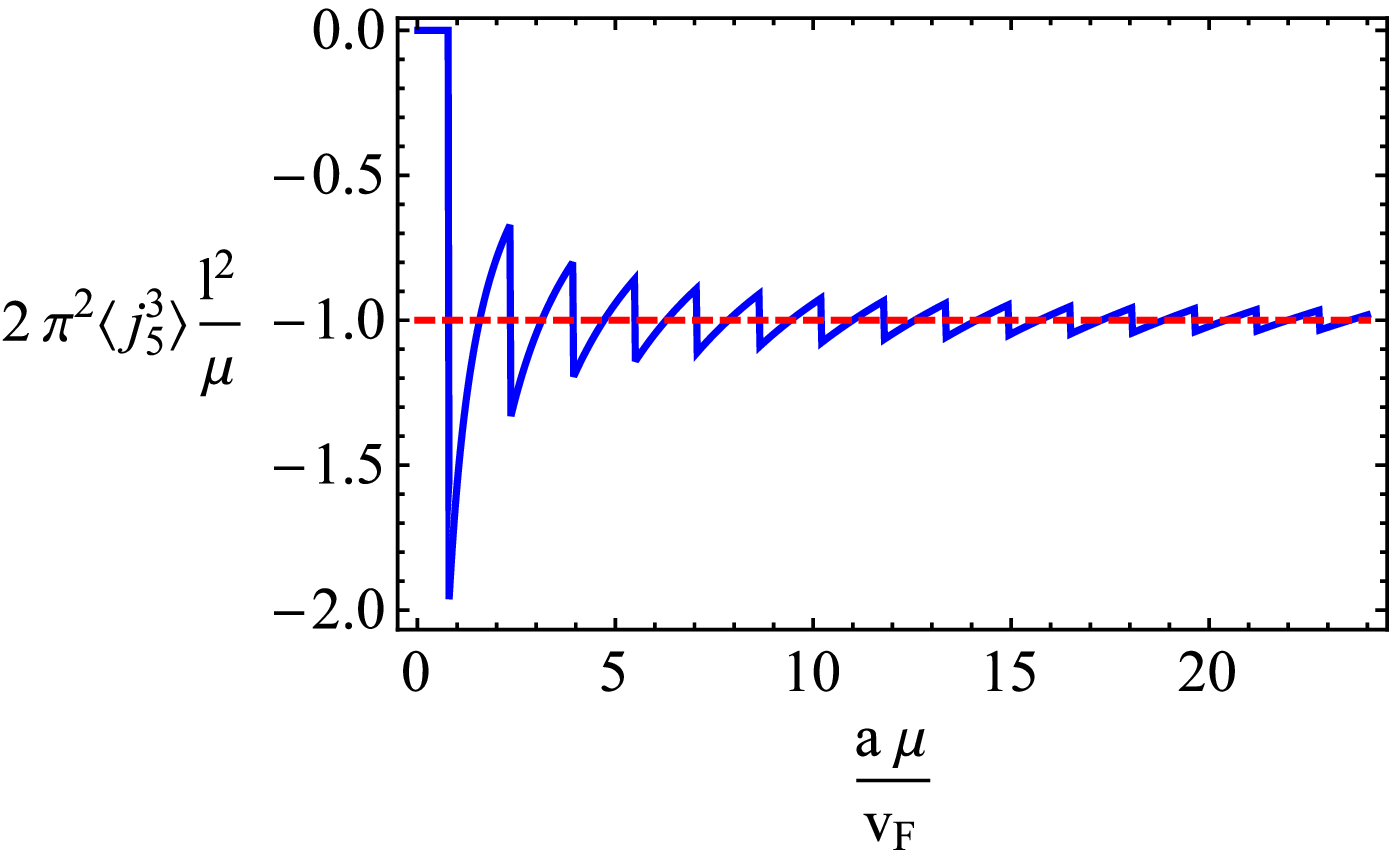}
\hfill
\includegraphics[width=0.32\textwidth]{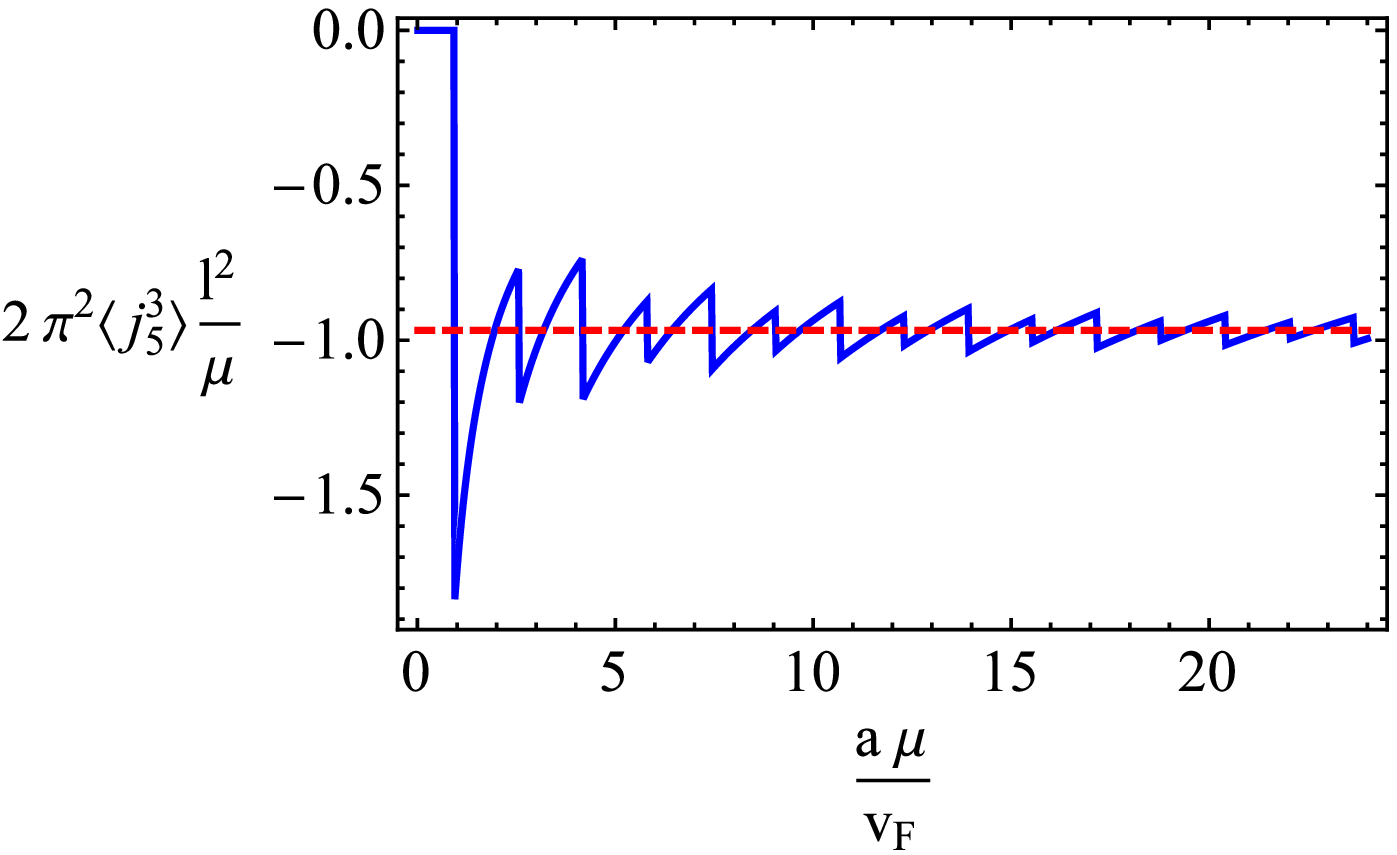}
\hfill
\includegraphics[width=0.32\textwidth]{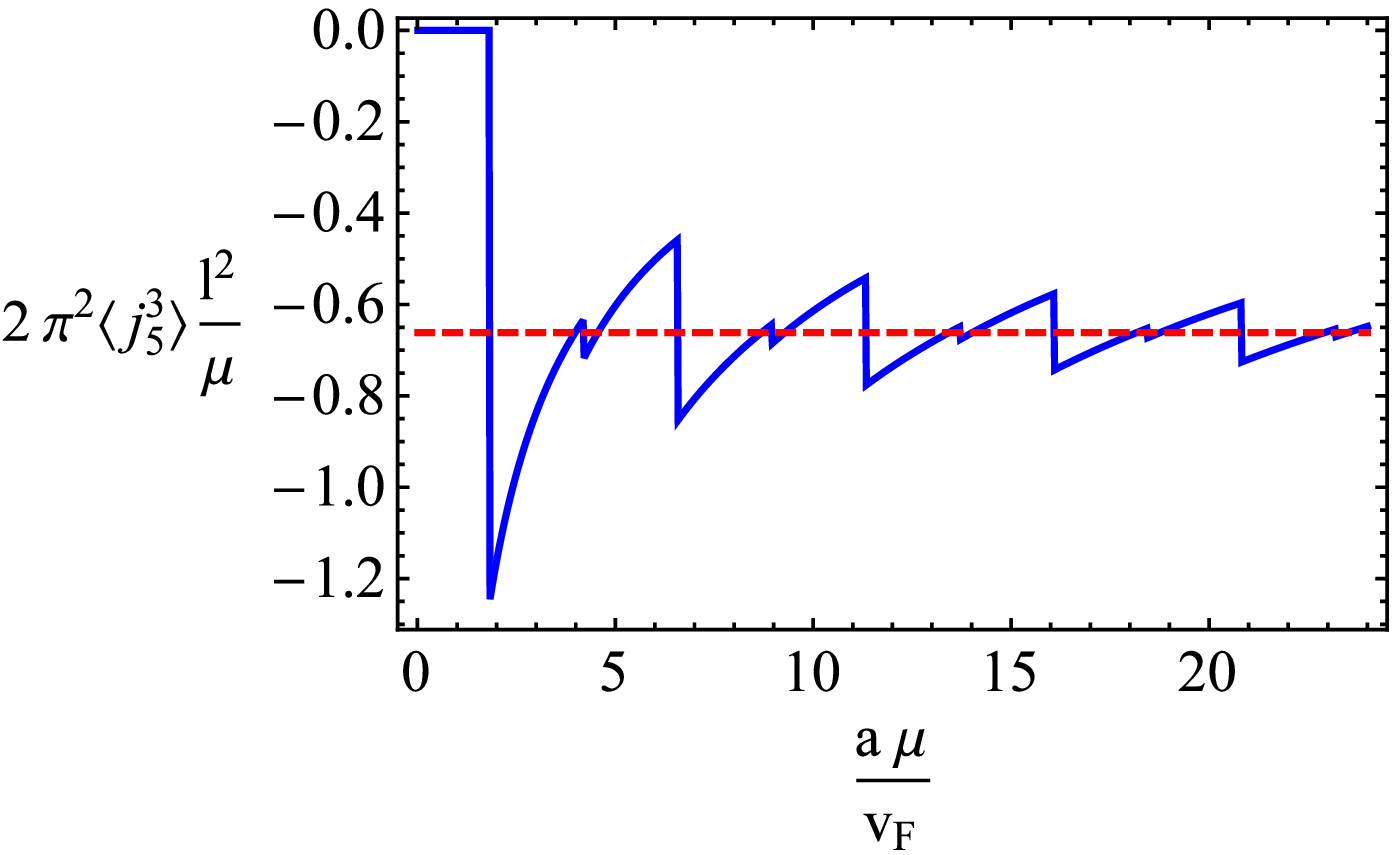}
\end{center}
\caption{(Color online) The dimensionless axial current density in an infinite space (red dashed line) and in the middle of
the slab (blue solid line) plotted as a function of the width of the slab $a$ for the three values of the band gap:
$m/\mu=0$ (left panel), $m/\mu=0.25$ (middle panel), and $m/\mu=0.75$ (right panel). To plot the figure, we 
fixed $v_F/\mu=12.5~\mbox{\AA}$.}
\label{fig:axial_current_slab_n0_a}
\end{figure*}

\begin{figure*}[ht]
\begin{center}
\includegraphics[width=0.32\textwidth]{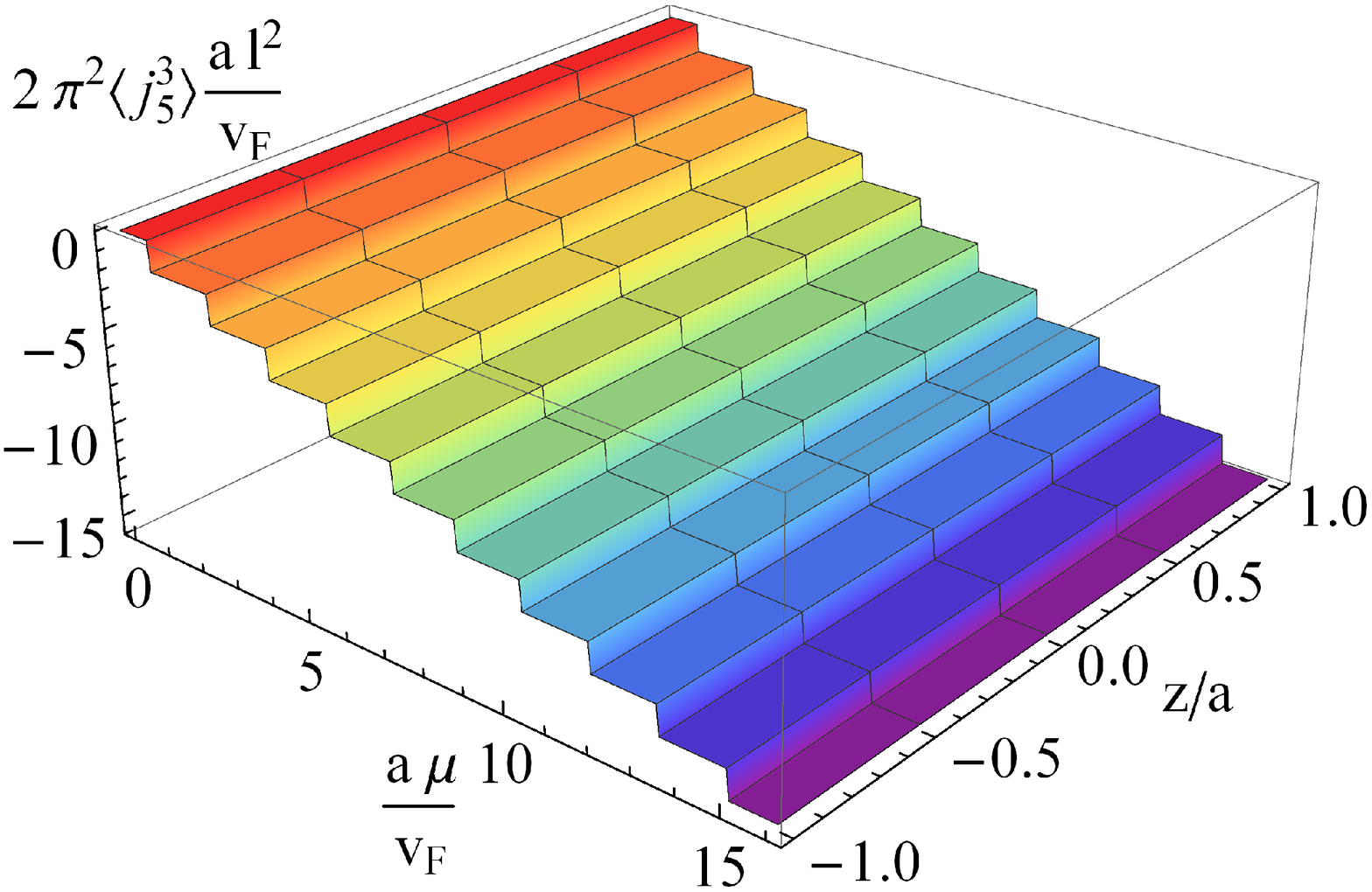}
\hfill
\includegraphics[width=0.32\textwidth]{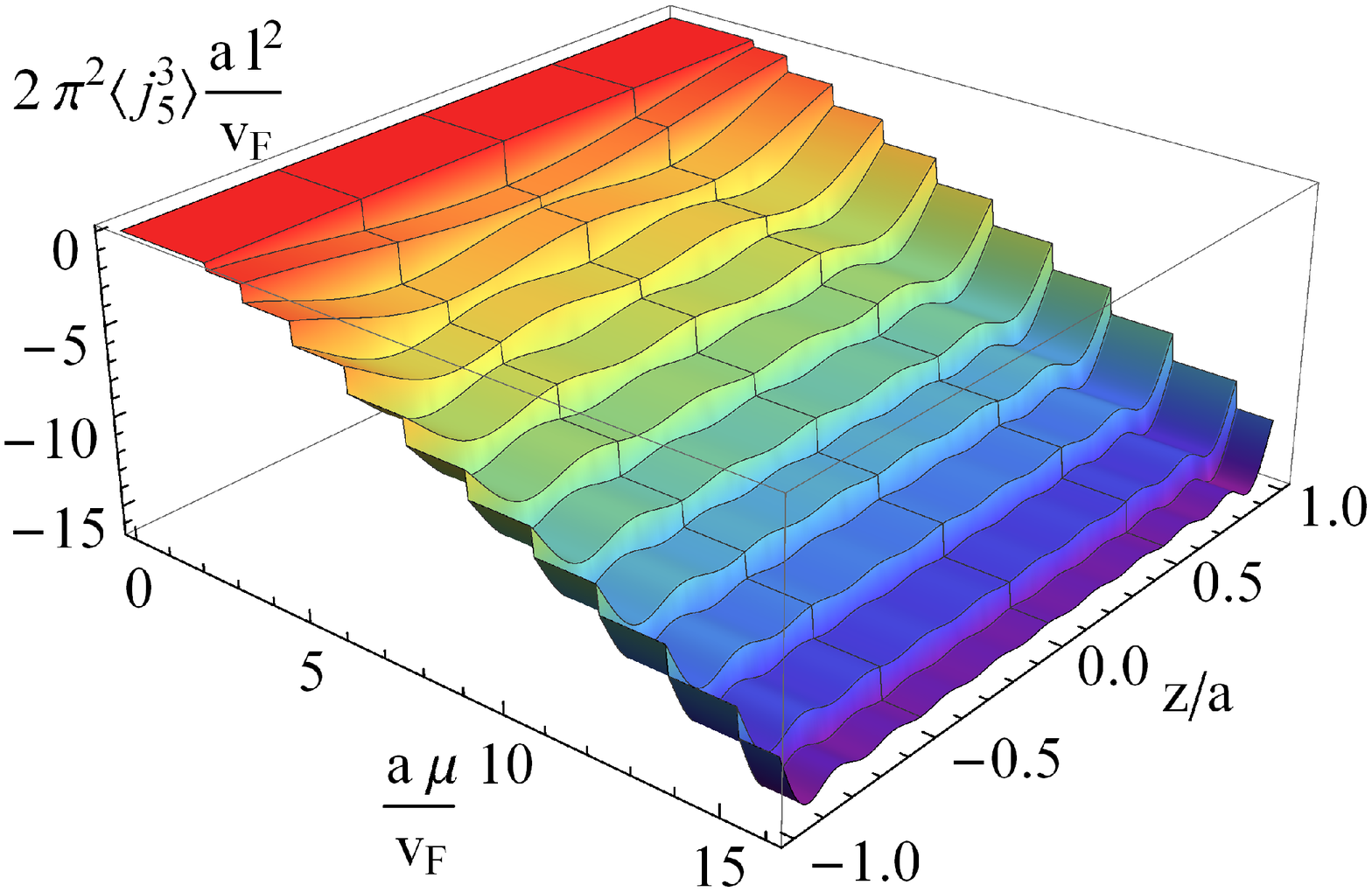}
\hfill
\includegraphics[width=0.32\textwidth]{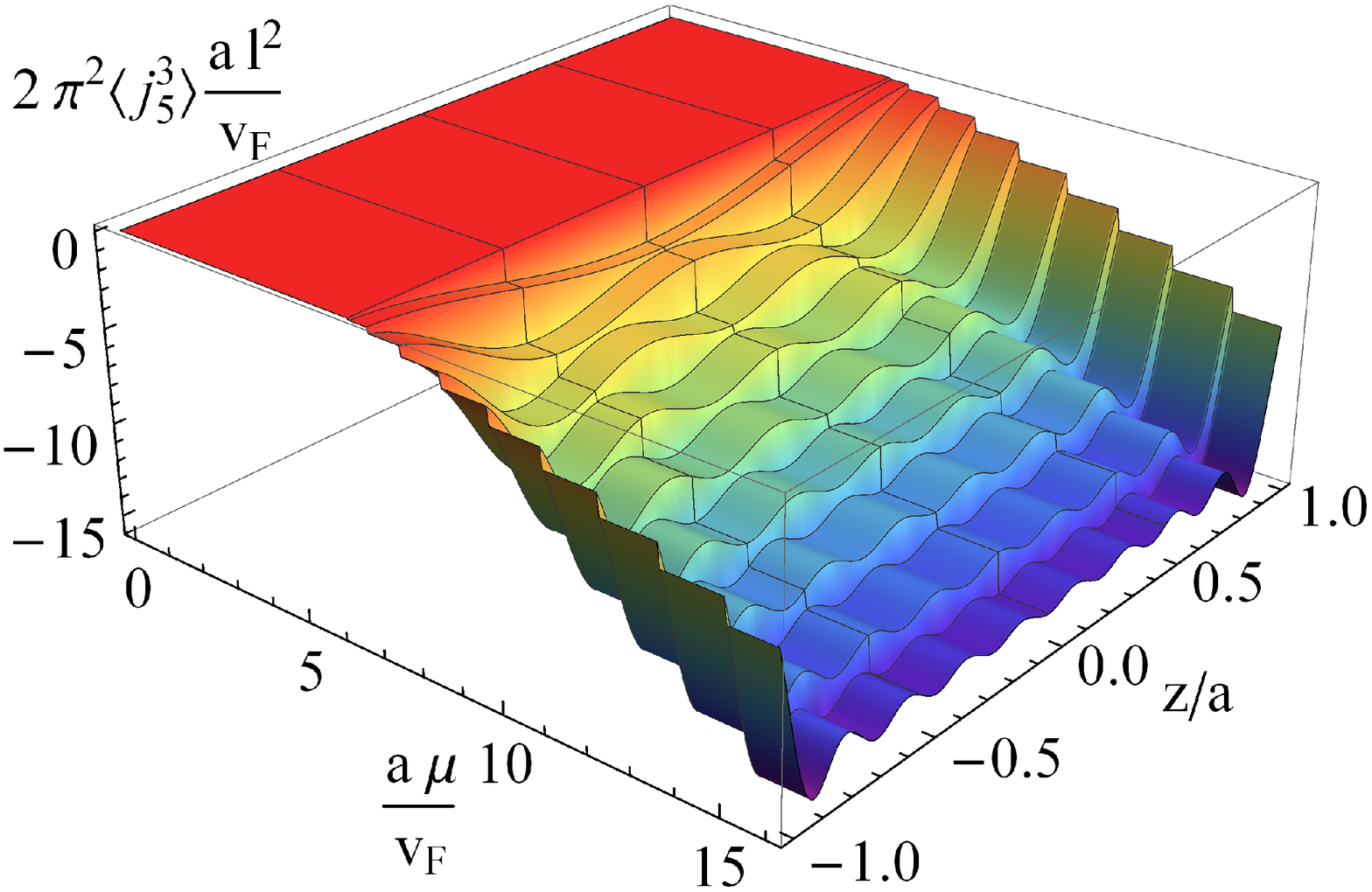}
\end{center}
\caption{(Color online) The dimensionless axial current density plotted as a function of the
chemical potential and $z/a$ for the three values of the band gap:
$am/v_F=0$ (left panel), $am/v_F=2$ (middle panel), and $am/v_F=6$ (right panel). 
To plot the figure, we fixed $v_F/a=25~\mbox{meV}$.}
\label{fig:axial_current_slab_3D}
\end{figure*}

Our numerical results for the axial current density (\ref{axial-current-slab-n0}) in a slab geometry
are presented in Figs.~\ref{fig:axial_current_slab_n0_mu}, \ref{fig:axial_current_slab_n0_a}, and 
\ref{fig:axial_current_slab_3D}. For comparison, in Figs.~\ref{fig:axial_current_slab_n0_mu} 
and \ref{fig:axial_current_slab_n0_a} we also show the results for an infinite space. The dependence of
the dimensionless axial current density $2\pi^2\langle j_{5}^{3}\rangle al^2/v_F$, in the middle of the
slab ($z=0$), as a function of the chemical potential is presented in Fig.~\ref{fig:axial_current_slab_n0_mu}.
[Recall that the axial current density in Eq.~(\ref{axial-current-slab-n0}) is not uniform
in space when $m\neq 0$.] The three panels show the results for three representative values of
the band gap: $am/v_F=0$ (left panel), $am/v_F=2$ (middle panel), and $am/v_F=6$ (right panel). 
To plot the figure, we used $v_F/a=25~\mbox{meV}$. As expected, a nonzero current density
is obtained only when $\mu>m$, i.e., when the chemical potential is larger than the size of the band gap.
In agreement with the earlier observation, we also see that the axial current density
is quantized in the slab.

The dependence of the dimensionless axial current density $2\pi^2\langle j_{5}^{3}\rangle l^2/\mu$, in the
middle of the slab ($z=0$), as a function of the width is shown in Fig.~\ref{fig:axial_current_slab_n0_a}.
The three panels show the results for the following three values of the band gap: $m/\mu=0$ (left panel),
$m/\mu=0.25$ (middle panel), and $m/\mu=0.75$ (right panel). To plot the figure, we used
$v_F/\mu=12.5~\mbox{\AA}$. As we find, the functional dependence of the current density on the
width of the slab has a sawtooth shape. This is rather natural consequence of the quantization of the
wave vector. When the size $a$ becomes large, the finite-size effects quickly decrease and the
result approaches the limit of an infinite system. We also see that a nonzero gap has a damping
effect in the dependence on $a$. [Notice the difference in the vertical scales in the three panels of
Fig.~\ref{fig:axial_current_slab_n0_a}.]

It is interesting to note that, in the gapped case ($m\neq 0$), the steps in the axial current
density have different heights, see Fig.~\ref{fig:axial_current_slab_n0_mu}. This is pronounced the most
in the lowest few steps of the current density. (Note, at the same time, that the widths of the steps in
$a\mu/v_F$ remain nearly, although not  exactly, the same.) The corresponding steps are determined
by the low-energy quasiparticle states with the smallest few wave vectors, i.e., the wave vectors
which are modified the most by a nonzero gap. This can be checked explicitly in the limit of a small
but nonzero band gap. In such a limit, an analytical expression for the wave vectors can be obtained
approximately by solving the spectral equation (\ref{spectrum-n0-matching-main}). The result reads
$p^{(m)}_{z,k} \simeq (2k-1) \pi /(4a) + 2m/[\pi v_F (2k-1)]$ where $k$ is a positive integer. By making
use of this, one can check that the contributions from the states with small $k$ vary a lot because
of great variations in the values of $\cos(2p^{(m)}_{z,k} a)$; see Eq.~(\ref{axial-current-slab-n0}).
(Away from the middle point in the slab, the heights of the steps also vary. This is clear
from Fig.~\ref{fig:axial_current_slab_3D}.) With increasing the value of $a|\mu|/v_F$, on the other
hand, the effect of the gap diminishes and the functional dependence of the axial current density
gradually approaches the result in the gapless limit. This is understandable since the states with
large wave vectors, which dominate the result in such a regime, are insensitive to the size of the gap.

As is clear from Eq.~(\ref{axial-current-slab-n0}), in the case of a nonzero band gap, the axial current density
is not uniform: it depends on the position $z$ inside the semimetal. The corresponding dependence is shown
explicitly in Fig.~\ref{fig:axial_current_slab_3D}, where we present the results for the gapless case $am/v_F=0$ 
(left panel) alongside with the results for the two cases with nonzero band gaps: $am/v_F=2$ (middle panel) 
and $am/v_F=6$ (right panel). The fact that the axial current density is constant
inside the slab in the gapless case is consistent with the continuity equation. Indeed, in the case
of the vanishing gap and in the absence of electric fields, the axial current is conserved inside
the slab. In contrast, there are clearly visible oscillations of the axial current density in the other two cases
when $m \ne 0$ (see the middle and right panels in Fig.~\ref{fig:axial_current_slab_3D}). Taking into
account that the axial current is not conserved in the gapped case, such oscillations are not forbidden.
Moreover, as we see, the larger is the gap, the more pronounced are the oscillations. When the gap
is fixed, we also find that the local amplitude of the axial current density oscillations is not the same
across the whole sample: it decreases with the increasing distance from the slab boundaries.

It may be interesting to emphasize that, in a finite-thickness slab, the axial current density originates
from helicity-correlated standing waves. Of course, the underlying roots of this are (i) finite thickness of
the slab and (ii) the spin-polarized nature of the LLL. This can be seen explicitly from the structure of
the LLL wave function in Eq.~(\ref{psi-matching-n0-slab-01}), which takes the following simple form
in the gapless limit:
\begin{eqnarray}
\Psi_{\rm slab}(\mathbf{r})_{n=0}= -  \frac{e^{ip_y y-\left( l p_y+x/l\right)^2/2} }{2\pi^{1/4} \sqrt{a l}}
\left(
 \begin{array}{c}
   0 \\
   ie^{2iap_z -i p_z z} \\
   0  \\
   e^{ip_z z} \\
 \end{array}
   \right) .
\label{psi-m0-n0-slab-text}
\end{eqnarray}
Such a standing wave is made of a pair of counterpropagating plane waves carrying opposite chiralities
or helicities. This is clear since, in the chiral representation used, see Eq.~(\ref{Dirac-matrices}), the
upper (lower) two components of the wave function describe right-handed (left-handed) modes.
The configuration in Eq.~(\ref{psi-m0-n0-slab-text}) can be interpreted as follows: the helicity of each
plane wave propagating to the boundary flips the sign after the reflection. This is indeed expected for
the spin-polarized LLL states. A nontrivial helicity correlation of the LLL standing waves remains
even in the gapped case, but it is not as transparent. Such a correlation is the key feature
that is responsible for the axial current density in a semimetal slab.

Before concluding this subsection, let us briefly discuss the implications of the chiral separation
effect in a slab of finite thickness. Naively, one may expect that the axial current density in the bulk of the
semimetal should lead to an accumulation of positive chiral charge on one side of the slab, and negative
chiral charge on the other. As is easy to check, however, this does not occur. In fact, the axial
charge density is identically zero everywhere, $\langle j_5^0 \rangle \equiv 0$.

\subsection{Axial current in the vacuum with finite band gap}
\label{App:BC-M}

As the results in the previous subsection show, in a slab geometry, there is a nonzero axial current density in the
bulk of semimetal. However, the current should be vanishing in the vacuum because of the imposed boundary
conditions. Then, the current density should presumably go to zero in the surface layer. In order to clarify this
issue, in this subsection, we investigate the continuity equation for the axial current density at the vacuum side
of the slab.

In the absence of an electric field, the axial charge and current densities should satisfy the following continuity
equation \cite{ABJ}:
\begin{equation}
\partial_{0}j_5^0 + \bm{\nabla} \cdot \mathbf{j}_5=2m(z)i\psi^{\dag}\gamma^0\gamma^5\psi,
\label{continuity-equation-axial-current-0}
\end{equation}
in all regions of space, including at the boundary of the slab. Note that, in order to have the
continuity equation well-defined, we will assume that the vacuum band gap $M$ is large,
but finite. The corresponding results for the LLL wave functions are derived in Appendix~\ref{App:BC-M},
see Eqs.~(\ref{psi-matching-n0-slab-01-Mvacuum}), (\ref{psi-matching-n0-Mvacuum-top-m}),
and (\ref{psi-matching-n0-Mvacuum-bottom-m}). Also, in the same appendix, we derive a
modified version of the spectral equation for the wave vector $p_z$; see
Eq.~(\ref{spectrum-n0-matching-Mvacuum}).

By making use of the wave functions in Eqs.~(\ref{psi-matching-n0-Mvacuum-top-m}) and
(\ref{psi-matching-n0-Mvacuum-bottom-m}), we can also compute the axial current density
outside the slab. In the region $z>a$, the corresponding result is given by
\begin{eqnarray}
\label{vacuum-current-m}
\langle j_{5}^{3}\rangle_{n=0, z>a} &=& -\frac{|eB|v_F\sign{(\mu)}}{2\pi a} \sum_{p_z}\theta\left(\mu^2-v_F^2p_z^2-m^2\right)   \nonumber \\
&\times&\frac{v_F^2p_z^2M e^{-2(z-a) \sqrt{M^2-v_F^2p_z^2-m^2}/v_F} }{2(v_F^2p_z^2+m^2)(M-m)+\frac{mv_F\sqrt{M^2-(v_F^2p_z^2+m^2)}}{a}+\frac{v_F^3p_z^2M}{a\sqrt{M^2-(v_F^2p_z^2+m^2)}}},
\end{eqnarray}
where the sum is performed over the discrete values of the wave vectors that satisfy
Eq.~(\ref{spectrum-n0-matching-Mvacuum}).
The result in the region $z<-a$ is similar, but one should replace $-(z-a)\to(z+a)$.

In order to check the continuity equation in the regions outside the slab, let us calculate the
ground-state expectation value of the quantity that appears on right-hand side of
Eq.~(\ref{continuity-equation-axial-current-0}). It is straightforward to show that
\begin{eqnarray}
\label{vacuum-RHS-m}
2iM\langle\psi_{z>a}^{\dag}\gamma^0\gamma^5\psi_{z>a}\rangle &=&\frac{|eB|\sign{(\mu)}}{2\pi a} \sum_{p_z}\theta\left(\mu^2-v_F^2p_z^2-m^2\right) 2\sqrt{M^2-v_F^2p_z^2-m^2}
 \nonumber \\
&\times& \frac{v_F^2p_z^2M e^{-2(z-a) \sqrt{M^2-v_F^2p_z^2-m^2}/v_F} }{2(v_F^2p_z^2+m^2)(M-m)+\frac{mv_F\sqrt{M^2-(v_F^2p_z^2+m^2)}}{a}+\frac{v_F^3p_z^2M}{a\sqrt{M^2-(v_F^2p_z^2+m^2)}}}.
\end{eqnarray}
In the region $z<-a$, the corresponding quantity is obtained by replacing $-(z-a)\to(z+a)$ and flipping
the overall sign.

By taking the derivative of Eq.~(\ref{vacuum-current-m}) with respect to $z$ and making use
of Eq.~(\ref{vacuum-RHS-m}), we check that the continuity equation (\ref{continuity-equation-axial-current-0})
is indeed satisfied in the regions outside the slab. Therefore, the axial current flows to the boundary,
where, due to the presence of the large vacuum mass $M$, it exponentially vanishes. Moreover,
as claimed earlier, this happens without any chiral charge accumulation at the boundaries.

\section{Chiral magnetic effect in the slab}
\label{sec:chiral-current-mu5}

In this section, we will study the chiral magnetic effect in a slab geometry (for simplicity, only
the case of the chiral limit will be considered here). In order to do this, we
need to introduce a nonzero chiral chemical potential $\mu_5$ into our model of a semimetal.
In the chiral limit, one might try to introduce $\mu_5$ by just replacing the fermion number chemical
potential $\mu$ with $\mu\pm\mu_5$ in the distribution function (\ref{distribution-T0}) for the
right- and left-handed particles, respectively. By recalling, however, that in the presence of the
boundaries the chiral symmetry is broken even in the case of gapless Dirac fermions in
the bulk, we know that $\mu_5$ does not correspond to a conserved quantity. In such a
situation, it is more convenient to treat $\mu_5$ as a phenomenological parameter
that modifies the model Hamiltonian, i.e.,
\begin{equation}
H= \int d^3 \mathbf{r} \, \Psi^\dagger (\mathbf{r})
\left[v_F \bm{\alpha} \cdot(-i\bm{\nabla}+e\mathbf{A})-\mu-\mu_5\gamma^5  \right]
\Psi (\mathbf{r}).
\label{hamiltonian-5}
\end{equation}
It is worth noting, that in the model under consideration $\mu_5$ determines the relative energy 
shift of the left- and right-handed Weyl nodes.
We calculate the ground-state expectation values by using the Schwinger prescription, where the
summation over energy eigenvalues is performed with the distribution function
$f_{0}(E)=-\mbox{sign}{(E)}/2$. [Note that, in this case, the effective distribution function
$f_{0}(E_{-}) + f_{0}(E_{+})$, where $E_{\pm}=-\mu\pm \sqrt{v_F^2p_z^2+2n\epsilon_L^2}$,
gives the same distribution as Eq.~(\ref{distribution-T0}) in the gapless limit.] 

For Hamiltonian (\ref{hamiltonian-5}), we find the following LLL wave function inside the semimetal:
\begin{equation}
\Psi_{\rm slab}(\mathbf{r})_{n=0}= Y_{0}(\xi)e^{ip_yy}\left\{ e^{ip_zz} C_0 \left(
 \begin{array}{c}
   0 \\
   1 \\
   0  \\
   0 \\
 \end{array}
   \right) +e^{i\tilde{p}_zz} \tilde{C}_0 \left(
 \begin{array}{c}
   0 \\
   0 \\
   0  \\
   1 \\
 \end{array}
   \right)\right\},
\label{psi-matching-n0-semimetal-5}
\end{equation}
where $p_z=-(E+\mu+\mu_5)/v_F$ and $\tilde{p}_z=(E+\mu-\mu_5)/v_F$. The corresponding wave functions outside the
slab are given by Eqs.~(\ref{psi-matching-n0-vacuum-top}) and (\ref{psi-matching-n0-vacuum-bot}). By making
use of the matching conditions in Eqs.~(\ref{psi-matching-boundary1}) and (\ref{psi-matching-boundary2}),
we derive the spectral equation for the wave vectors, $\cos{[a(p_z-\tilde{p}_z)]}=0$. By solving this equation,
we find that the energy parameter $E$ can take only the following discrete values:
\begin{equation}
E_k=-\mu+\frac{\pi v_F}{2a} \left(k-\frac{1}{2}\right),
\label{spectrum-n0-5}
\end{equation}
where $k$ is a positive integer. After enforcing the boundary conditions, we also derive the following
final expression for the wave function inside the slab:
\begin{equation}
\Psi_{\rm slab}(\mathbf{r})_{n=0}=  \frac{Y_{0}(\xi)e^{ip_yy}}{2\sqrt{al}}  \left(
 \begin{array}{c}
   0 \\
   e^{ip_zz} \\
   0  \\
   -ie^{i p_z a+i\tilde{p}_z(z-a)} \\
 \end{array}
   \right),
\label{psi-matching-n0-semimetal-matched-5}
\end{equation}
and find that the corresponding LLL contribution to the electric current vanishes,
\begin{eqnarray}
\langle j^{3}\rangle=\sum_{k=1}^{\infty} f(E_k)v_F\Psi_{\rm slab}^{\dag}(\mathbf{r})_{n=0}
\gamma^0\gamma^3\Psi_{\rm slab}(\mathbf{r})_{n=0} = 0.
\label{charge-current-n0-slab-5}
\end{eqnarray}
Although this result is natural in the Bogolyubov model, it may appear surprising
because it is precisely the LLL contribution that saturates the CME in an infinite system.
To complete the analysis of the CME in a
finite slab, however, we should still analyze the contributions of higher Landau levels ($n>0$).
For a semimetal described by Hamiltonian (\ref{hamiltonian-5}), the wave functions in higher
Landau levels are given by
\begin{eqnarray}
\Psi_{\rm slab}(\mathbf{r})_{n}&=&e^{ip_yy}\Vast[ C_1e^{iP_zz}\left(
         \begin{array}{c}
           -i v_F \left(\sqrt{P_z^2+ \frac{2n\epsilon_{L}^2}{v_F^2}} +P_z\right)Y_{n-1}(\xi) \\
           \sqrt{2n\epsilon_{L}^2}Y_{n}(\xi) \\
           0 \\
           0 \\
         \end{array}
       \right) +C_2e^{i\tilde{P}_zz}\left(
         \begin{array}{c}
           0 \\
           0 \\
           iv_F \left(\sqrt{\tilde{P}_z^2+\frac{2n\epsilon_{L}^2}{v_F^2}}-\tilde{P}_z\right)Y_{n-1}(\xi)
           \\
           \sqrt{2n\epsilon_{L}^2}Y_{n}(\xi) \\
         \end{array}
       \right) \nonumber \\
       &+& C_3e^{-iP_zz}\left(
         \begin{array}{c}
           -iv_F \left(\sqrt{P_z^2+\frac{2n\epsilon_{L}^2}{v_F^2}}-P_z\right)Y_{n-1}(\xi) \\
           \sqrt{2n\epsilon_{L}^2}Y_{n}(\xi) \\
           0 \\
           0 \\
         \end{array}
       \right) +C_4e^{-i\tilde{P}_zz}\left(
         \begin{array}{c}
           0 \\
           0 \\
           iv_F \left(\sqrt{\tilde{P}_z^2+\frac{2n\epsilon_{L}^2}{v_F^2}}+\tilde{P}_z\right)Y_{n-1}(\xi)
           \\
           \sqrt{2n\epsilon_{L}^2}Y_{n}(\xi) \\
         \end{array}
       \right)
 \Vast], \nonumber\\
\label{phi-explicit-5}
\end{eqnarray}
where $P_z=v_F^{-1}\sqrt{(E+\mu+\mu_5)^2-2n\epsilon_{L}^2}$ and
$\tilde{P}_z=v_F^{-1}\sqrt{(E+\mu-\mu_5)^2-2n\epsilon_{L}^2}$.

The corresponding spectral equation is obtained by matching the wave functions in the bulk with
those in the vacuum; see Eqs.~(\ref{psi-matching-boundary1}) and (\ref{psi-matching-boundary2}).
Its explicit form reads
\begin{equation}
\left(1-e^{4iaP_z}\right)  \left(1-e^{4ia\tilde{P}_z}\right)
\left[(E+\mu)^2-\mu_5^2-2n\epsilon_{L}^2\right]
+v_F^2P_z\tilde{P}_z\Big(1+e^{4iaP_z}
+e^{4ia\tilde{P}_z} + 4e^{2ia\left(P_z+\tilde{P}_z\right)} + e^{4ia\left(P_z+\tilde{P}_z\right)}\Big) =0 .
\label{spectrum-n-5}
\end{equation}
When $\mu_5\neq0$, we could solve this equation only numerically. By making use of the spectral equation,
we can write down the formal solution for the semimetal wave function in the slab. The corresponding expression
is bulky and not very informative. Therefore, we do not present it here. Instead, by making use of the numerical
solutions to Eq.~(\ref{spectrum-n-5}), we calculate the contribution of the higher Landau levels to the electric
and axial current densities. Both results vanish within the numerical precision used.

Thus, unlike the axial current density of the CSE, the electric current density of the CME is absent in the slab.
This result might have been expected from general considerations.
Because of the Bogolyubov bag model boundary conditions (with an infinite band gap in vacuum), there should
be no electric current across the boundary. Taking into account that the electric current is not anomalous, this
means that the continuity equation for the electric current can be satisfied only if the electric current vanishes
inside the slab. This is exactly what the direct calculations give. Moreover, in the model at hand, the
electric current will remain vanishing even in the limit $a\to \infty$. This fact is consistent with the case of 
infinite systems in equilibrium considered in Refs.~\cite{Zhou, Basar, Landsteiner}.

In passing, let us briefly discuss the simplest case of a Weyl semimetal. The corresponding model
Hamiltonian will be similar to that in Eq.~(\ref{hamiltonian-5}), but will include an additional term,
$-v_F \int d^3 \mathbf{r} \,\Psi^\dagger (\mathbf{r}) \left(\bm{\alpha} \cdot \mathbf{\Delta} \right) \gamma^5\Psi (\mathbf{r})$.
The value of $\mathbf{\Delta} \equiv \Delta\mathbf{e}_z$, which is often called the chiral shift parameter, determines
the separation between the Weyl nodes in momentum space. Here, for simplicity, we assume that
$\mathbf{\Delta}$ points in the $+z$ direction. This may be sufficient because the components of the chiral
shift parallel to the slab surface are not expected to modify the currents in the slab. When
$\Delta\neq 0$, we find that the LLL wave functions in the bulk have the same form as in
Eq.~(\ref{psi-matching-n0-semimetal-5}), but with $p_z=-(E+\mu+\mu_5)/v_F+\Delta$ and
$\tilde{p}_z=(E+\mu-\mu_5)/v_F-\Delta$. Then, by making use of the wave functions outside
the slab, see Eqs.~(\ref{psi-matching-n0-vacuum-top}) and (\ref{psi-matching-n0-vacuum-bot}),
and enforcing the boundary conditions in Eqs.~(\ref{psi-matching-boundary1}) and (\ref{psi-matching-boundary2}),
we find that the energy can take only the following discrete values:
\begin{equation}
E_k=-\mu+v_F\Delta+\frac{\pi v_F}{2a} \left(k-\frac{1}{2}\right),
\label{spectrum-n0-b}
\end{equation}
where $k$ is a positive integer. Compared to the result in Eq.~(\ref{spectrum-n0-5}), the only
difference here is the change: $\mu\to\mu-v_F\Delta$. This finding is in agreement with the result
in an infinite system, obtained in Ref.~\cite{chiral-shift-1}. The calculation of the higher Landau
level contributions to the axial current is more challenging in the general case $\mu_5\neq0$.
Therefore, we restrict our discussion to the simpler case, $\mu_5=0$. In contrast to the case of
an infinite system, we find that the contributions of higher Landau levels to the axial current
density vanish. Therefore, our results for the
axial and electric current densities in a Dirac semimetal slab, see Eqs.~(\ref{axial-current-slab-n00})
and (\ref{charge-current-n0-slab-5}), remain qualitatively the same also in the case of a Weyl
semimetal slab, but with the replacement $\mu\to\mu-v_F\Delta$.

\section{Work function, electric fields, and continuity equation for axial current}
\label{sec:electric-field}

The work function of a solid is defined as the energy needed to remove an electron from the solid to
vacuum (see, e.g., \cite{Venables}). Microscopically, it can be thought of as the result of a ``confining"
electric field $\bm{\mathcal{E}}$ near the surface, resulting from the electron density leaking slightly
out of the material. In the problem at hand, we consider a semimetal slab with a nonzero magnetic field
perpendicular to its surface. This means that $\bm{\mathcal{E}}\cdot\mathbf{B} \ne 0$ near the surfaces.
As we know, such a field configuration has a nontrivial (anomalous) contribution to the continuity equation
for the axial current \cite{ABJ}, i.e.,
\begin{equation}
\partial_{0} j_5^0  + \bm{\nabla} \cdot \mathbf{j}_5=2im(z)\psi^{\dag}\gamma^0\gamma^5\psi
-\frac{e^2}{4\pi^2}\bm{\mathcal{E}}\cdot\mathbf{B}.
\label{continuity-equation-axial-current}
\end{equation}
Therefore, it is important to investigate the role of this relation near the surfaces.

Let us assume that the electric field $\bm{\mathcal{E}}$ is perpendicular to the surface
and exists only in a thin layer of depth $\lambda_E$. We can conveniently describe
such a configuration by using a time-like component of the vector potential, $A_0$, i.e.,
\begin{eqnarray}
-a-\lambda_E<z<-a: & \quad &A_0=-\mathcal{E}z, \\
a<z<a+\lambda_E: & \quad &A_0=\mathcal{E}z.
\label{EF-definition-VacEF}
\end{eqnarray}
By assumption, $A_0$ vanishes inside the slab (i.e., $-a<z<a$), as well as in the vacuum regions
outside the thin surface layers (i.e., for $z<-a-\lambda_E$ and $z> a+\lambda_E$).

For the sake of clarity and brevity, we consider the case of zero gap inside the semimetal, $m=0$.
Inside the semimetal, where electric field is absent, the LLL wave function is given by
\begin{eqnarray}
\psi_{-a<z<a}  = e^{-iEt} Y_0(\xi)e^{ip_yy} \left(
                          \begin{array}{c}
                            0 \\
                            C_1e^{-iz p_z} \\
                            0 \\
                            C_2e^{iz p_z} \\
                          \end{array}
                        \right).
\label{wave-function-semimet-Vac1}
\end{eqnarray}
Although the Dirac equation admits exact analytic solutions in constant electric and magnetic
fields \cite{Bagrov}, we find it more convenient and transparent to obtain the corresponding
solutions in the region near the slab boundaries, where electric field is present, in the first
order of the perturbation theory in electric field $\mathcal{E}$. Since we treat electric field
in perturbation theory and only the LLL modes contribute to the axial current in a magnetic
field, we begin with the following ansatz [compare with the LLL wave functions
(\ref{wave-function-semimet-Vac1}) in the absence of electric field]:
\begin{eqnarray}
\psi_{\rm vac}=e^{-iEt}Y_0(\xi)e^{ip_yy} \left(
                          \begin{array}{c}
                            0 \\
                            \phi_1(z) \\
                            0 \\
                            \phi_2(z) \\
                          \end{array}
                        \right).
\label{wave-function-VacEF}
\end{eqnarray}
Then the Dirac equation in the region $a<z<a+\lambda_E$ implies the following equations
for the functions $\phi_1$ and $\phi_2$:
\begin{eqnarray}
E\phi_1(z) -v_F (i\partial_z+ e\mathcal{E}z/v_F )\phi_1(z)+M\phi_2(z) &=& 0, \\
E\phi_2(z) +v_F (i\partial_z-e\mathcal{E}z/v_F)\phi_2(z)+M\phi_1(z) &=& 0.
\label{wave-function-eq-NewEF}
\end{eqnarray}
We seek solutions of this system of equations as the Taylor series in $\mathcal{E}$ retaining
only its two first terms
\begin{eqnarray}
\phi_1(z)&=&f_0(z)-e\mathcal{E}f_1(z), \\
\phi_2(z)&=&g_0(z)-e\mathcal{E}g_1(z).
\label{wave-function-eq-2-VacEF}
\end{eqnarray}
By substituting Eq.~(\ref{wave-function-eq-2-VacEF}) into Eq.~(\ref{wave-function-eq-NewEF}),
we obtain the solutions
\begin{eqnarray}
f_0(z) &=& A_1e^{-p_z^{\prime}z}+A_2e^{p_z^{\prime}z}, \quad g_0(z) = -\frac{1}{M}\left[A_1e^{-p_z^{\prime}z}
(E+iv_Fp_z^{\prime})+A_2e^{p_z^{\prime}z}(E-iv_Fp_z^{\prime})\right],\\
f_1(z) &=& \frac{1}{4(v_Fp_z^{\prime})^3}\left\{A_1e^{-p_z^{\prime}z}
\left[i(v_F^2p_z^{\prime}+2M^2z)-2E^3z^2/v_F+E(v_F+2v_Fp_z^{\prime}z+2(Mz)^2/v_F)-2iE^2z\right]+\right. \nonumber \\
&&\left.+A_2e^{p_z^{\prime}z}\left[i(v_F^2p_z^{\prime}-2M^2z)+2E^3z^2/v_F-E(v_F-2v_Fp_z^{\prime}z+2(Mz)^2/v_F)+2iE^2z\right]\right\},\\
g_1(z) &=& -\frac{1}{4(v_Fp_z^{\prime})^3M}\left\{A_1e^{-p_z^{\prime}z}\left[-2E^3z^2 (E+iv_Fp_z^{\prime})/v_F
+M^2(v_F+2v_Fp_z^{\prime}z+2Ez^2(E+iv_Fp_z^{\prime})/v_F)\right]+\right. \nonumber \\
&&\left.+A_2e^{p_z^{\prime}z}\left[2E^3z^2(E-iv_F p_z^{\prime})/v_F-M^2(v_F-2v_Fp_z^{\prime}z+2Ez^2(E-iv_Fp_z^{\prime})/v_F)\right]\right\},
\label{wave-function-eq-2-exp-VacEF}
\end{eqnarray}
where $p_z^{\prime}=v_F^{-1}\sqrt{M^2-E^2}$ and $A_1$ and $A_2$ are constants. By matching the wave function
(\ref{wave-function-semimet-Vac1}) and the vacuum functions
\begin{eqnarray}
\psi_{z<-a-\lambda_E} &=& e^{-iEt} Y_0(\xi)e^{ip_yy} e^{p_z^{\prime}(z+a+\lambda_E)}C^{\prime \prime}_1 \left(
                                 \begin{array}{c}
                                   0 \\
                                   1 \\
                                   0 \\
                                   i \\
                                 \end{array}
                               \right),
                               \\
\psi_{-a-\lambda_E<z<-a}  &=& e^{-iEt} Y_0(\xi)e^{ip_yy}  \left(
                          \begin{array}{c}
                            0 \\
                            f_0(z+a)+e\mathcal{E}f_1(z+a) \\
                            0 \\
                            g_0(z+a)+e\mathcal{E}g_1(z+a) \\
                          \end{array}
                        \right),
\label{bottom-region}
                        \\
\psi_{a<z<a+\lambda_E}  &=& e^{-iEt} Y_0(\xi)e^{ip_yy}  \left(
                          \begin{array}{c}
                            0 \\
                            f_0(z-a)-e\mathcal{E}f_1(z-a) \\
                            0 \\
                            g_0(z-a)-e\mathcal{E}g_1(z-a) \\
                          \end{array}
                        \right),
\label{top-region}
                        \\
\psi_{z>a+\lambda_E} &=& e^{-iEt} Y_0(\xi)e^{ip_yy}  e^{-p_z^{\prime}(z-a-\lambda_E)}C^{\prime}_1\left(
                                 \begin{array}{c}
                                   0 \\
                                   1 \\
                                   0 \\
                                   -i \\
                                 \end{array}
                               \right),
    \label{wave-function-semimet-VacEF}
\end{eqnarray}
at $z=-a-\lambda_E$, $z=-a$, $z=a-\lambda_E$, and $z=a$, and taking into account
the normalization conditions, we find that the linear in $\mathcal{E}$ terms of the wave
functions (\ref{bottom-region}) and (\ref{top-region}) do not contribute to the continuity
equation for axial current because
\begin{eqnarray}
&&\partial_{3} \langle j_{5}^{3}\rangle_{-a-\lambda_E<z<-a} -2iM \langle\psi^{\dag}_{-a-\lambda_E<z<-a}\gamma^0\gamma^5
\psi^{\dag}_{-a-\lambda_E<z<-a}\rangle = \mathcal{O}\big(e^2\mathcal{E}^2\big), \\
&&\partial_{3} \langle j_{5}^{3}\rangle_{a<z<a+\lambda_E}
-2iM\langle\psi^{\dag}_{a<z<a+\lambda_E}\gamma^0\gamma^5\psi^{\dag}_{a<z<a+\lambda_E}\rangle= \mathcal{O}\big(e^2\mathcal{E}^2\big).
\label{derivative-current-VacEF}
\end{eqnarray}
This result is quite natural because the anomalous term in the chiral anomaly is connected
with the lack of a chiral symmetry invariant regularization of the famous linearly divergent
triangle diagram \cite{ABJ}. Thus, it is high-energy modes whose contribution is divergent
and should be regularized which are responsible for the anomalous term in the continuity
equation for axial current. Consequently, we conclude that the low-energy modes confined
in the semimetal satisfy the continuity equation for axial current without the anomalous term.

\section{Conclusion}
\label{sec:Conclusion}

In this paper, we studied the chiral separation and chiral magnetic effects in a Dirac semimetal
with a slab geometry placed in a constant magnetic field perpendicular to its surfaces. We used
the Bogolyubov boundary conditions \cite{Bogoliubov} at the surfaces of the slab. It is worth
mentioning that such a model was originally used in high-energy physics for the description of
hadrons within the framework of the bag models of quarks \cite{Thomas}. This model assumes
that the quasiparticles have a large band gap (Dirac mass) in the vacuum regions
outside the slab.

Using this model setup, we derived analytically the spectral equation and the wave functions for
the bulk modes. Furthermore, we calculated the axial current density and found that, just like in
an infinite system, only the LLL modes contribute to this quantity. We show that the main
consequence of a finite slab thickness is that the axial current density becomes quantized.
The corresponding quantization could be revealed in its functional dependence on the chemical
potential and the thickness of the slab. The underlying reason for the quantization is the discretization
of the wave vectors and energy levels in the slab. In other words, the axial current density originates
from helicity-correlated standing waves, associated with the LLL. It is also interesting to point
that there is no chiral charge accumulation on the semimetal boundaries, as might have been
naively expected. The size of the quantization steps in the axial current density as a function
of the chemical potential, see Fig.~\ref{fig:axial_current_slab_n0_mu},
is given by $\delta\langle j_{5}^{3}\rangle= eB v_F /(4a\pi)$. Because the steps are inversely
proportional to the thickness of the slab $a$, most likely the corresponding structure can be
observed in experiment only if the samples are sufficiently thin.

As we show in this study, the dependence of the axial current density on the thickness of the
slab $a$ has a very characteristic sawtooth shape, see Fig.~\ref{fig:axial_current_slab_n0_a}.
With increasing the value of $a$, the quantization effects become less pronounced and the axial
current density gradually approaches the limit of an infinite system. Interestingly, the quantization
persists even in the case of Dirac quasiparticles with nonzero gaps. However, a nonzero
gap has a damping effect in the dependence on $a$.

We also find that, in the case of gapped Dirac fermions, $m\neq0$, the axial current density
is not uniform across the semimetal slab, but oscillates as a function of the position,
see Fig.~\ref{fig:axial_current_slab_3D}. This is
in contrast to the chiral limit when the axial current is constant. In fact, the larger is the gap,
the more pronounced are the oscillations. Formally, such a behavior is connected with a
nonconservation of the axial charge. It appears that the amplitude of the oscillations decreases
with increasing the distance from the boundaries. To the best of our knowledge, the possibility
of such oscillations and their key features have not been reported before. It would be very
interesting to test this scenario in the lattice simulations. Also, this phenomenon could potentially
be investigated experimentally in a semimetal with a spontaneously generated gap, e.g., via
magnetic catalysis \cite{MCatalysis}, or another mechanism.

By introducing a nonzero chiral chemical potential $\mu_5$, we also analyzed the chiral magnetic
effect in a semimetal slab. In contrast to the axial current density due to the CSE, the corresponding
electric current due to the CME is absent in a semimetal slab. In retrospect, the vanishing result is
a natural outcome of the non-anomalous continuity equation for the electric current and the Bogolyubov
boundary conditions \cite{Thomas}. Indeed, the latter are equivalent to the requirement of
vanishing electric current from the semimetal to vacuum. It is worth emphasizing, however,
that in this study we limited our consideration only to static configurations. This automatically
excludes all transient phenomena in which nonzero electric currents are generated out of equilibrium
and evolve in time. The corresponding generation of the CME in finite-size samples would be very
interesting and should be studied in the future. Such a problem is beyond the scope of this
study however.

We took into account a nonzero electric field near the slab surface created by a double charge layer
present at the surfaces of solids. It was found that, because of the presence of this electric field,
the continuity equation for the axial current is fulfilled for the low-energy modes considered in this
study without taking into account the linear in electric field anomalous term. In all fairness, the
current understanding of the chiral anomaly in the problem of a finite-size Dirac semimetal remains
incomplete. While it is understood that the chiral symmetry is explicitly broken by the surface
effects, it is not completely clear whether the model implementation of the boundary layer used
in the present study is sufficient to capture all relevant physics effects.

One of the limitations of this study is the use of a special orientation of the magnetic field
perpendicular to the slab surfaces. The main reason for this was that the analysis would become
much more challenging in the case of a tilted magnetic field. From the physics viewpoint, the
complications come from the need of squeezing the Landau orbits into a finite thickness of
the slab. Nevertheless, here we could speculate that, in the case of a general orientation
of the magnetic field, one may consider separately the perpendicular and parallel
(with respect to the surface) components of the currents, while the results for the perpendicular
components of the currents are expected to remain qualitatively the same as in the special case
considered in this paper (except for the replacement $B\to B\cos{\theta}$, where $\theta$ is
an angle between the direction of the magnetic field and the normal to the slab surface).
The parallel components of the currents are most likely to be the same as in an unbounded
infinite system, but with $B\to B\sin{\theta}$. While plausible, such a scenario clearly requires
a further in-depth investigation, which is beyond the scope of this study.

\acknowledgments

The work of E.V.G. was supported partially by the Ukrainian State Foundation for Fundamental
Research. The work of V.A.M. was supported by the Natural Sciences and Engineering
Research Council of Canada. The work of I.A.S. was supported in part by the
U.S. National Science Foundation under Grant No.~PHY-1404232.

\appendix

\section{Derivation of the Landau-level wave functions}

\label{App:WF}

In this appendix we derive the Landau-level wave functions in the model described by the
Hamiltonian in Eq.~(\ref{hamiltonian-0}). We look for the solutions of the Dirac equation,
$H\psi(\mathbf{r})=E\psi(\mathbf{r})$, in the form $\psi(\mathbf{r})=e^{ip_zz+ip_yy}\phi(x)$.
In this case, function $\phi(x)$ satisfies the following equation:
\begin{equation}
\left[\left(-v_Fp_z\gamma^z-v_Fp_y\gamma^y+iv_F\gamma^x\partial_x\right)\gamma^0-v_FeBx\gamma^y\gamma^0+m\gamma^0\right]\phi(x)=E\phi(x).
\label{equation-01}
\end{equation}
In the case of a constant magnetic field in the $+z$ direction, assumed here, we choose the
vector potential in the Landau gauge, $\mathbf{A}=(0, Bx, 0)$. (Clearly, all observables should
be independent of a specific gauge choice.) Instead of the $x$ coordinate, it is convenient to
introduce a new dimensionless variable,
\begin{equation}
\xi=\sqrt{|eB|}\left(\frac{p_y}{eB}+x\right) ,
\label{notation-01}
\end{equation}
and rewrite Eq.~(\ref{equation-01}) as follows:
\begin{equation}
\left[\partial_{\xi}+is_{\perp}\xi\gamma^y\gamma^x-\frac{i\gamma^x}{\sqrt{|eB|}}\left(p_z\gamma^z+(m/v_F)-(E/v_F)\gamma^0\right)\right]\phi(\xi)=0,
\label{equation-02}
\end{equation}
where $s_{\perp}=\sign{(eB)}$. By making use of the same approach as in Ref.~\cite{wf}, let us introduce
a set of linearly independent bispinors $u^{\pm}_{s}$ that satisfy the following relations:
\begin{eqnarray}
i\gamma^y\gamma^x u^{\pm}_{s} &=& \mp u^{\pm}_{s}, \label{bispinors1}\\
\gamma^y\gamma^x\left(E\gamma^0-v_Fp_z\gamma^z\right)u^{\pm}_{s} &=& s\sqrt{v_F^2p_z^2-E^2}\, u^{\pm}_{s},
\label{bispinors2}\\
\frac{\gamma^x\left(\gamma^0E-\gamma^zv_Fp_z-m\right)}{\sqrt{E^2-v_F^2p_z^2-m^2}} \, u^{-}_{s} &=& u^{+}_{s}  .
\label{bispinors3}
\end{eqnarray}
Then, by expressing the wave function $\phi(\xi)$ as a linear combination of the two bispinors,
\begin{equation}
\phi_{s}(\xi) = \Phi^{s}_{+}(\xi)u^{+}_{s}+\Phi^{s}_{-}(\xi)u^{-}_{s},
\label{phi-bispinors}
\end{equation}
and using the relations in Eqs.~(\ref{bispinors1}) through (\ref{bispinors3}), we rewrite the equation for the
wave function (\ref{equation-02}) in the following form:
\begin{equation}
\left[\partial_{\xi}\Phi_{\pm}^{s}(\xi)\mp s_{\perp}\xi\Phi_{\pm}^{s}(\xi)+i\kappa\Phi_{\mp}^{s}(\xi)\right]=0, \qquad
\left[\partial_{\xi}^2 \mp s_{\perp}-\xi^2+\kappa^2\right]\Phi_{\pm}^{s}(\xi)=0,
\label{equation-04}
\end{equation}
where $\kappa= \sqrt{E^2-v_F^2p_z^2-m^2}/\epsilon_L $ and $\epsilon_L\equiv v_F\sqrt{|eB|}$.
Without loss of generality, let us choose $s_{\perp}\equiv\sign{(eB)}=+1$. Then, the solutions of
Eq.~(\ref{equation-04}) are given in terms of the parabolic cylinder functions \cite{Bateman}:
\begin{equation}
\Phi_{-}^{s}(\xi) = D_{\kappa^2/2}(\sqrt{2}\xi), \qquad \Phi_{+}^{s}(\xi) = \frac{i\kappa}{\sqrt{2}}D_{\kappa^2/2-1}(\sqrt{2}\xi).
\label{solutions-01}
\end{equation}
By requiring that the wave functions are finite at $|\xi|\rightarrow\infty$, one finds that $\kappa^2/2=n$ where
$n=0, 1, 2 \ldots$ are nonnegative integers. In this special case, the parabolic cylinder functions
$D_{n}(\sqrt{2} \xi)$ can be expressed in terms of the Hermite polynomials $H_{n}(\xi)$:
$D_{n}(\sqrt{2}\xi) = \frac{1}{\sqrt{2^n}}e^{-\xi^2/2}H_{n}(\xi)$. By making use of the definition of $\kappa$,
we also obtain the corresponding Landau-level energies: $E_n^2=v_F^2p_z^2+m^2+2n\epsilon_L^2$.
It should be noted that function $\Phi_{+}^{s}(\xi)$ is not finite at $|\xi|\rightarrow\infty$ when $n=0$.
However, by using the relation in Eq.~(\ref{bispinors3}), we can express $u^{+}_{s}$ through $u^{-}_s$, i.e.,
\begin{equation}
u^{+}_s=\frac{-m\gamma^x-\gamma^ys\sqrt{v_F^2p_z^2-E_n^2}}{\sqrt{E_n^2-v_F^2p_z^2-m^2}}u^{-}_s = -\gamma^x\frac{m+s\sqrt{E_n^2-v_F^2p_z^2}}{\sqrt{E_n^2-v_F^2p_z^2-m^2}}u^{-}_s.
\label{um-up}
\end{equation}
Then, in the case of the LLL ($n=0$), we see that the coefficient in the last equation vanishes if $s=-1$.
This means that, for the LLL, only one value of the spin, $s=-1$, is allowed. For the higher Landau levels,
both spin projections, i.e., $s=\pm1$, are allowed. The Landau-level wave functions are given by
\begin{equation}
\phi_s(\xi)=\left[ \frac{e^{-\xi^2/2}}{\sqrt{2^n}}H_{n}(\xi)-i\gamma^x \frac{m+s\sqrt{m^2+2n\epsilon_L^2}}{\sqrt{2\epsilon_L^2}} \frac{e^{-\xi^2/2}}{\sqrt{2^{n-1}}}H_{n-1}(\xi) \right]u^{-}_s.
\label{phi-explicit}
\end{equation}
where, according to Eqs.~(\ref{bispinors1}) through (\ref{bispinors3}),
\begin{equation}
u^{-}_s =\left(
                                                         \begin{array}{c}
                                                           0 \\
                                                           \chi_2 \\
                                                           0 \\
                                                           \chi_4 \\
                                                         \end{array}
                                                       \right), \qquad u^{+}_s = \frac{m+s\sqrt{E_n^2-v_F^2p_z^2}}{\sqrt{E_n^2-v_F^2p_z^2-m^2}} \left(
                                                         \begin{array}{c}
                                                           -\chi_4 \\
                                                           0 \\
                                                           \chi_2 \\
                                                           0 \\
                                                         \end{array}
                                                       \right), \qquad \chi_4= \frac{s\sqrt{E_n^2-v_F^2p_z^2}}{E_n-v_Fp_z}\chi_2.
\label{um-equation-01}
\end{equation}
Finally, the explicit form of the Landau-level wave functions reads
\begin{eqnarray}
\label{psi-explicit-n0}
\psi(\mathbf{r})_{n=0}&=&C_0\,e^{ip_zz+ip_yy}\,\phi_{0; s=-1}(\xi),\\
\label{psi-explicit-n}
\psi(\mathbf{r})_{n\neq0}&=&e^{ip_zz+ip_yy} \left[C_1\phi_{n; s=+1}(\xi)+C_2\phi_{n; s=-1}(\xi)\right].
\end{eqnarray}
where
\begin{eqnarray}
\label{phi-explicit-n-s+}
\phi(\xi)_{n>0, s=+1}&=& C_1 \left(
                                                  \begin{array}{c}
                                                    -i\frac{\sqrt{m^2+2n\epsilon_L^2}}{E_n-v_Fp_z} \frac{m+\sqrt{m^2+2n\epsilon_L^2}}{\sqrt{2n\epsilon_L^2}} Y_{n-1}(\xi) \\
                                                     Y_{n}(\xi) \\
                                                     i\frac{m+\sqrt{m^2+2n\epsilon_L^2}}{\sqrt{2n\epsilon_L^2}} Y_{n-1}(\xi)  \\
                                                     \frac{\sqrt{m^2+2n\epsilon_L^2}}{E_n-v_Fp_z}Y_{n}(\xi) \\
                                                  \end{array}
                                                \right),  \\
\label{phi-explicit-n-s-}
\phi(\xi)_{n>0, s=-1}&=& C_2 \left(
                                                  \begin{array}{c}
                                                    i\frac{\sqrt{m^2+2n\epsilon_L^2}}{E_n-v_Fp_z} \frac{m-\sqrt{m^2+2n\epsilon_L^2}}{\sqrt{2n\epsilon_L^2}} Y_{n-1}(\xi) \\
                                                     Y_{n}(\xi) \\
                                                     i\frac{m-\sqrt{m^2+2n\epsilon_L^2}}{\sqrt{2n\epsilon_L^2}} Y_{n-1}(\xi)  \\
                                                     -\frac{\sqrt{m^2+2n\epsilon_L^2}}{E_n-v_Fp_z}Y_{n}(\xi) \\
                                                  \end{array}
                                                \right), \\
\label{phi-explicit-n0-s-}
\phi(\xi)_{n=0, s=-1}&=& C_0 \left(
                                                  \begin{array}{c}
                                                    0 \\
                                                     Y_{0}(\xi) \\
                                                     0  \\
                                                     -\frac{m}{E_0-v_Fp_z}Y_{0}(\xi) \\
                                                  \end{array}
                                                \right).
\end{eqnarray}
Here $Y_n(\xi)=\frac{e^{-\xi^2/2}}{\sqrt{2^n n!\sqrt{\pi}} }H_n(\xi)$ are the harmonic oscillator wave functions.
By making use of Eqs.~(\ref{psi-explicit-n0}) through (\ref{phi-explicit-n0-s-}), it is straightforward to obtain the
wave functions (\ref{LLL}) and (\ref{higher}) in the main text.

\section{Matching solutions at the semimetal boundaries}
\label{App:BC}

In this appendix, we derive the explicit expressions for the wave functions in the semimetal with a slab
geometry by matching the general solutions, presented in Eqs.~(\ref{LLL-slab}) and (\ref{higher-slab})
in the main text, with the corresponding vacuum solutions. This is done by making use of the Bogolyubov
bag model, in which the wave functions outside the semimetal satisfy the same type Dirac equation,
but with a large vacuum band gap $M$.

\subsection{Matching the wave functions of the $n=0$ modes}
\label{sec:AC-slab-infmass}

Let us first consider the $n=0$ modes. In order to obtain normalizable solutions outside the semimetal, in the
regions with $z>a$ and $z<-a$, in Eq.~(\ref{LLL}) we will replace $p_z$ with $ip_z^{\prime}$ and $-ip_z^{\prime}$,
respectively. With such a parametrization of the vacuum solutions, the energy of the $n=0$ modes is given by
$E = \pm\sqrt{M^2-(v_F p_z^{\prime})^2}$. In the limit $M \to \infty$, the corresponding solutions must have the same
(finite) energy as the $n=0$ modes in the bulk. This implies that $v_F p_z^{\prime} \approx M-(m^2+v_F^2p_z^2)/(2M)$.
Therefore, to leading order in the inverse powers of $M$, the corresponding wave functions in the vacuum are
given by
\begin{eqnarray}
\label{psi-matching-n0-vacuum-top}
\psi(\mathbf{r})_{n=0, z>a} &=& C_0^{\prime}Y_{0}(\xi)e^{ip_yy} e^{-M(z-a)/v_F }  \left(
 \begin{array}{c}
   0 \\
   1 \\
   0  \\
   -i \\
 \end{array}
   \right),  \\
\label{psi-matching-n0-vacuum-bot}
\psi(\mathbf{r})_{n=0, z<-a} &=& C_0^{\prime \prime}Y_{0}(\xi)e^{ip_yy} e^{M(z+a)/v_F } \left(
 \begin{array}{c}
   0 \\
   1 \\
   0  \\
   i \\
 \end{array}
   \right).
\end{eqnarray}
By making use of Eqs.~(\ref{LLL-slab}), (\ref{psi-matching-n0-vacuum-top}), and (\ref{psi-matching-n0-vacuum-bot}),
we find that the matching conditions on the two sides of the semimetal slab,
\begin{eqnarray}
\Psi_{\rm slab}(x,y,a)_{n=0} &=&\psi(x,y,a)_{n=0, z>a},
\label{psi-matching-boundary1} \\
\Psi_{\rm slab}(x,y,-a)_{n=0}&=&\psi(x,y,-a)_{n=0, z<-a} ,
\label{psi-matching-boundary2}
\end{eqnarray}
lead to the following equation for $p_z$:
\begin{equation}
v_F p_z\cos{(2ap_z)}+m\sin{(2ap_z)}=0,
\label{spectrum-n0-matching}
\end{equation}
where $p_z\neq0$. Note that $p_z=0$ is not allowed because it corresponds to a trivial solution
for the wave function.  Satisfying the boundary conditions also fixes the value of constant $\tilde{C}_0$ in
the superposition of two counterpropagating waves in the bulk solution (\ref{LLL-slab}):
\begin{equation}
\tilde{C}_0=-e^{2ip_za}C_0 \frac{E_0\left(E_0+v_Fp_z\right)}{m\left(m-i v_F p_z\right)}.
\label{tC0-n0-matching}
\end{equation}
By taking this into account, we derive the following final expression for the wave function
inside the semimetal:
\begin{eqnarray}
\Psi_{\rm slab}(\mathbf{r})_{n=0}= C_0 Y_{0}(\xi)e^{ip_yy} \left(
 \begin{array}{c}
   0 \\
   2e^{iap_z}\frac{v_Fp_z\cos\left[p_z(z-a)\right]-(m+iE_0)\sin\left[p_z(z-a)\right]}{im+v_Fp_z -E_0} \\
   0  \\
   -2ie^{iap_z}\frac{v_Fp_z\cos\left[p_z(z-a)\right]-(m-iE_0)\sin\left[p_z(z-a)\right]}{im+v_Fp_z -E_0} \\
 \end{array}
   \right) ,
\label{psi-matching-n0-slab-01}
\end{eqnarray}
where
\begin{equation}
|C_0|^2=\frac{ \epsilon_{L }\left[m^2+v_Fp_z(v_Fp_z-E_0)\right]}
{4 v_F \left[2a(m^2+v_F^2p_z^2)+mv_F\right]}
\end{equation}
is obtained from the condition of the wave function normalization.

\subsection{Matching the wave functions of $n>0$ modes}

Let us now consider the modes  with $n > 0$ outside the slab. As in the case of the $n=0$ mode,
in order to obtain normalizable solutions in the $z>a$ and $z<-a$ regions, in Eq.~(\ref{higher})
we replace $p_z$ with $ip_z^{\prime}$ and $-ip_z^{\prime}$, respectively. The energies of such solutions
are given by $E_n=\pm \sqrt{M^2+2n\epsilon_{L}^2-(v_Fp_z^{\prime})^2}$. In the limit $M\to \infty$,
these should coincide with the corresponding expressions for the Landau-level energies in the bulk.
This is satisfied if we choose $v_Fp_z^{\prime}\approx M-(m^2+v_F^2p_z^2)/(2M)$ in the vacuum solutions.
By taking this into account, we derive the following vacuum wave function in the $z>a$ region:
\begin{equation}
\label{psi-matching-n-vacuum-top-00}
\psi(\mathbf{r})_{n, z>a} =
e^{-p_z^{\prime}(z-a)+ip_yy} \left[ M \left(C_{+}^{\prime}+C_{-}^{\prime}\right) \frac{Y_{n-1}(\xi)}{\sqrt{2n\epsilon_{L}^2}} \left(
 \begin{array}{c}
    1 \\
    0 \\
    i \\
    0 \\
 \end{array}
   \right) + \left(
 \begin{array}{c}
    -iE_n\left(C_{+}^{\prime}+C_{-}^{\prime}\right) \frac{Y_{n-1}(\xi)}{\sqrt{2n\epsilon_{L}^2}} \\
    C_{+}^{\prime}Y_{n}(\xi)  \\
    0  \\
    iC_{-}^{\prime}Y_{n}(\xi) \\
 \end{array}
   \right) + \mathcal{O}\left(\frac{1}{M}\right)\right],
\end{equation}
where we kept terms up to subleading order in the inverse powers of $M$. From the normalization of
the wave function (\ref{psi-matching-n-vacuum-top-00}), we find that
\begin{equation}
C_{\pm}^{\prime} = C_{1}^{\prime} \frac{\sqrt{2n\epsilon_{L}^2}}{2M} \pm  C_{2}^{\prime} +\mathcal{O}\left(\frac{1}{M^2}\right).
\label{vacuum-coefficients-bot}
\end{equation}
Therefore, the final expression for the wave function in the vacuum region $z>a$ is given by
\begin{equation}
\label{psi-matching-n-vacuum-top-01}
\psi(\mathbf{r})_{n, z>a} = e^{-M(z-a)/v_F+ip_yy} \left[ C_1^{\prime} Y_{n-1}(\xi) \left(
      \begin{array}{c}
         1\\
         0 \\
         i \\
         0 \\
      \end{array}
    \right) + C_2^{\prime} Y_{n}(\xi)  \left(
      \begin{array}{c}
        0 \\
        1 \\
        0 \\
        -i \\
      \end{array}
    \right) \right] .
\end{equation}
The wave function in the other vacuum region, $z<-a$, can be obtained in a similar way. The final result reads
\begin{equation}
\label{psi-matching-n-vacuum-bottom-01}
\psi(\mathbf{r})_{n, z<-a} = e^{M(z+a)/v_F+ip_yy} \left[ C_1^{\prime \prime} Y_{n-1}(\xi) \left(
      \begin{array}{c}
        -1 \\
         0 \\
         i \\
         0 \\
      \end{array}
    \right) + C_2^{\prime \prime}  Y_{n}(\xi) \left(
      \begin{array}{c}
        0 \\
        1 \\
        0  \\
        i  \\
      \end{array}
    \right) \right].
\end{equation}
The matching conditions at the boundary of the semimetal are similar to those in Eqs.~(\ref{psi-matching-boundary1})
and (\ref{psi-matching-boundary2}), i.e.,
\begin{eqnarray}
\Psi_{\rm slab}(x,y,a)_{n > 0} &=&\psi(x,y,a)_{n > 0, z>a},
\label{psi-matching-boundary-ng00}
\\
\Psi_{\rm slab}(x,y,-a)_{n > 0}&=&\psi(x,y,-a)_{n > 0, z<-a}.
\label{psi-matching-boundary-ng0}
\end{eqnarray}
By substituting the wave functions from Eqs.~(\ref{higher-slab}), (\ref{psi-matching-n-vacuum-top-01}),
and (\ref{psi-matching-n-vacuum-bottom-01}) into these matching conditions, we find that $p_z$ should
satisfy the same spectral equation (\ref{spectrum-n0-matching}) as in the case of the $n=0$ modes.
By taking into account, however, that the rank of the system of Eqs.~(\ref{psi-matching-boundary-ng00})
and (\ref{psi-matching-boundary-ng0}) is $2$ units less than the dimension of the system, we obtain the
following two linearly independent solutions for each higher Landau level ($n>0$) inside the slab:
\begin{eqnarray}
\Psi^{(1)}_{\rm slab}(\mathbf{r})_{n > 0} &=& C_{+}e^{ip_yy} \left(
 \begin{array}{c}
   -2i\frac{\sqrt{2n\epsilon_{L}^2} e^{-ip_za}\sin\left[p_z(z+a)\right]}{m+i\left(v_Fp_z-E_n\right)} Y_{n-1}(\xi) \\
   2ie^{-iap_z}\frac{v_Fp_z\cos\left[p_z(z+a)\right]+\left(m-iE_n\right)\sin\left[p_z(z+a)\right]}{m+i\left(v_Fp_z-E_n\right)}Y_{n}(\xi) \\
   \frac{\sqrt{2n\epsilon_{L}^2} e^{-ip_za}\left[m+i(v_Fp_z+E_n)\right]\sin\left[p_z(z+a)\right]}{m^2+i v_F p_z m+2n\epsilon_{L}^2} Y_{n-1}(\xi)  \\
   -2e^{-iap_z}\frac{v_Fp_z\cos\left[p_z(z+a)\right]+(m+iE_n)\sin\left[p_z(z+a)\right]}{m+i\left(v_Fp_z-E_n\right)}Y_{n}(\xi) \\
 \end{array}
   \right),
   \label{psi-n-all-InfMass1} \\
\Psi^{(2)}_{\rm slab}(\mathbf{r})_{ n > 0} &=& C_{-}e^{ip_yy} \left(
 \begin{array}{c}
   2e^{-iap_z}\frac{i\left[m^2-im\left(v_Fp_z+E_n\right)\right] \sin\left[p_z(z+a)\right] +2n\epsilon_{L}^2
   \cos\left[p_z(z+a)\right] +v_F p_z\left[v_Fp_z+E_n\right]e^{ip_z(z+a)} }{\sqrt{2n\epsilon_{L}^2}\sqrt{m^2+2n\epsilon_{L}^2}}
   Y_{n-1}(\xi) \\
   -2\frac{e^{-ip_za} \left(v_Fp_z+E_n\right) \sin\left[p_z(z+a)\right]}{\sqrt{m^2+2n\epsilon_{L}^2}} Y_{n}(\xi) \\
  2i\frac{e^{-iap_z}\sqrt{m^2+2n\epsilon_{L}^2}\left[p_z\cos\left[p_z(z+a)\right]+\left(m-
  iE_n\right)\sin\left[p_z(z+a)\right]\right]}{\sqrt{2n\epsilon_{L}^2}\left(v_Fp_z-E_n\right)} Y_{n-1}(\xi)  \\
   2ie^{-iap_z}\frac{\left(v_Fp_z+E_n\right)\sin\left[p_z(z+a)\right]}{\sqrt{m^2+2n\epsilon_{L}^2}}Y_{n}(\xi) \\
 \end{array}
   \right), \nonumber\\
\label{psi-n-all-InfMass2}
\end{eqnarray}
where the normalization constants are given by
\begin{eqnarray}
|C_{+}|^2 &=&\frac{ \epsilon_{L} p_z\left[m^2+ n\epsilon_{L}^2+v_Fp_z\left(v_Fp_z-E_n\right)\right]}
{4v_F \left[ 2ap_zE_n^2+mv_Fp_z-n\epsilon_{L}^2\sin{(4ap_z)}\right]}, \\
|C_{-}|^2 &=& \frac{n \epsilon_{L}^{3} p_z\left(v_Fp_z-E_n\right)^2}{4 v_F (m^2+2n\epsilon_{L}^2)
\left[2ap_zE_n^2 +mv_Fp_z -n\epsilon_{L}^2\sin{(4ap_z)}\right]}.
\end{eqnarray}

\subsection{Matching of the LLL wave functions in the case of finite vacuum gap}

Let us now consider the LLL in the case of large but finite vacuum gap $M$. In order to derive the tails of the
LLL wave functions in the regions $z>a$ and $z<-a$ outside the slab, we use Eq.~(\ref{LLL}) with $p_z$
replaced with $ip_z^{\prime}$ and $-ip_z^{\prime}$, respectively, i.e.,
\begin{eqnarray}
\label{psi-matching-n0-Mvacuum-top}
\psi(\mathbf{r})_{n=0, z>a}&=&Y_{0}(\xi)e^{ip_yy} e^{-p_z^{\prime}(z-a)} A_0^{\prime} \left(
 \begin{array}{c}
   0 \\
   1 \\
   0  \\
   -\frac{M}{E_0-iv_Fp_z^{\prime}} \\
 \end{array}
   \right), \\
\label{psi-matching-n0-Mvacuum-bottom}
\psi(\mathbf{r})_{n=0, z<-a}&=&Y_{0}(\xi)e^{ip_yy} e^{p_z^{\prime}(z+a)} A_0^{\prime \prime} \left(
 \begin{array}{c}
   0 \\
   1 \\
   0  \\
   -\frac{M}{E_0+iv_Fp_z^{\prime}} \\
 \end{array}
   \right).
\end{eqnarray}
Here $p^{\prime}_z=v_F^{-1}\sqrt{M^2-E^2_0}$. Enforcing the matching conditions at the boundaries
of the slab, see Eqs.~(\ref{psi-matching-boundary1}) and (\ref{psi-matching-boundary2}), we find that
the wave vector $p_z$ should satisfy the following spectral equation:
\begin{equation}
\frac{-v_Fp_z\sqrt{M^2-E_0^2}\cos{(2ap_z)}+(E_0^2-mM)\sin{(2ap_z)}}{M}=0.
\label{spectrum-n0-matching-Mvacuum}
\end{equation}
[As expected, in the limit $M \to \infty$, this equation reduces to Eq.~(\ref{spectrum-n0-matching}).]
The corresponding LLL wave functions inside the semimetal and outside the slab are given by
\begin{eqnarray}
\label{psi-matching-n0-slab-01-Mvacuum}
\Psi_{\rm slab}(\mathbf{r})_{n=0}&=&A_0 Y_{0}(\xi)e^{ip_yy} \left(
 \begin{array}{c}
   0 \\
   e^{ip_z(z-a)} - e^{-ip_z(z-a)}\frac{(E_0+v_Fp_z)\left[E_0^2-mM -iv_Fp_z\sqrt{M^2-
   E_0^2}\right]}{mE_0(m-M)} \\
   0  \\
   \frac{m}{v_Fp_z-E_0}\left[e^{ip_z(z-a)} - e^{-ip_z(z-a)}\frac{(E_0-v_Fp_z)\left[E_0^2-mM -
   iv_Fp_z\sqrt{M^2-E_0^2}\right]}{mE_0(m-M)}\right] \\
 \end{array}
   \right),\\
\label{psi-matching-n0-Mvacuum-top-m}
\psi(\mathbf{r})_{n=0, z>a}&=&-A_0Y_{0}(\xi)e^{ip_yy} e^{-\sqrt{M^2-E_0^2}(z-a)/v_F} \frac{v_Fp_z\left[m(m-M) +(v_Fp_z+E_0)(v_Fp_z-i\sqrt{M^2-E_0^2})\right]}{mE_0(m-M)} \nonumber\\
&\times&\left(
 \begin{array}{c}
   0 \\
   1 \\
   0  \\
   -\frac{M}{E_0-i\sqrt{M^2-E_0^2}} \\
 \end{array}
   \right),\\
\label{psi-matching-n0-Mvacuum-bottom-m}
\psi(\mathbf{r})_{n=0, z<-a}&=&-2A_0Y_{0}(\xi)e^{ip_yy} e^{\sqrt{M^2-E_0^2}(z+a)/v_F} \frac{E_0m\left[v_Fp_z\cos{(2ap_z)} +\sqrt{M^2-E_0^2}\sin{(2ap_z)}\right]}{(E_0-v_Fp_z)\left[m(E_0-i\sqrt{M^2-E_0^2})-M(v_Fp_z+E_0)\right]} \nonumber\\
&\times&\left(
 \begin{array}{c}
   0 \\
   1 \\
   0  \\
   -\frac{M}{E_0+i\sqrt{M^2-E_0^2}} \\
 \end{array}
   \right).
\end{eqnarray}
where the overall constant $A_0$ is obtained from the normalization condition.
Its explicit expression is given by
\begin{equation}
|A_0|^2=\frac{\epsilon_L (m-M)E_0(E_0-v_Fp_z)}{4v_F\left(2aE_0^2(m-M)-mv_F\sqrt{M^2-E_0^2}-\frac{v_F^3p_z^2M}{\sqrt{M^2-E_0^2}}\right)}.
\end{equation}

\end{document}